\documentclass{IOS-Book-Article}

\usepackage{cite}

\usepackage[detect-mode=true,per-mode=symbol,binary-units=true]{siunitx}
\usepackage{multirow}

\newcolumntype{M}{>{\centering\arraybackslash}m}
\newcolumntype{L}{>{\raggedright\arraybackslash}m}

\usepackage{booktabs}
\usepackage[super]{nth}

\usepackage{graphicx}

\usepackage[normalsize]{subfigure}

\usepackage{csquotes} 

\usepackage[hidelinks]{hyperref}

\usepackage{enumitem}

\def\hb{\hbox to 10.7 cm{}}

\usepackage{wrapfig}

\usepackage{makeidx}
\makeindex

\begin{document}

\pagestyle{headings}
\def\thepage{}

\begin{frontmatter}              

\title{Fog Computing on Constrained Devices: Paving the Way for the Future IoT}

\markboth{}{June 2019\hb}

\author[A,B]{\fnms{Fl{\'a}via} \snm{Pisani}%
	\thanks{Corresponding Author: Av. Albert Einstein, 1251,
		Cidade Universit\'aria, Campinas/SP, Brazil, 13083-852; E-mail: \href{mailto:fpisani@ic.unicamp.br}{fpisani@ic.unicamp.br}.}},
\author[A]{\fnms{Fab{\'i}ola} \snm{M. C. de Oliveira}},
\author[A]{\fnms{Eduardo} \snm{S. Gama}},
\author[A,C]{\fnms{Roger} \snm{Immich}},
\author[A]{\fnms{Luiz F.} \snm{Bittencourt}},
and
\author[A]{\fnms{Edson} \snm{Borin}}

\runningauthor{F. Pisani et al.}
\address[A]{Institute of Computing, University of Campinas (UNICAMP), Brazil}
\address[B]{Department of Informatics, Pontifical Catholic University of\\Rio de Janeiro (PUC-Rio), Brazil}
\address[C]{Digital Metropolis Institute, Federal University of\\Rio Grande do Norte (UFRN), Brazil}

\begin{abstract}
In the long term, the Internet of Things~(IoT) is expected to become an integral part of people's daily lives.
In light of this technological advancement, an ever-growing number of objects with limited hardware may become connected to the Internet. 
In this chapter, we explore the importance of these constrained devices as well as how we can use them in conjunction with fog computing to change the future of the IoT.
First, we present an overview of the concepts of constrained devices, IoT, and fog and mist computing, and then we present a classification of applications according to the amount of resources they require (e.g., processing power and memory).
After that, we tie in these topics with a discussion of what can be expected in a future where constrained devices and fog computing are used to push the IoT to new limits.
Lastly, we discuss some challenges and opportunities that these technologies may bring.
\end{abstract}

\end{frontmatter}
\markboth{June 2019\hb}{June 2019\hb}

\section{Introduction}
Current prospects for the Internet of Things (IoT) indicate that, within the next few years, a global network of objects will connect tens of billions of devices through the Internet~\cite{Vaquero2014-FindingYourWayInTheFog}.
Most of these devices will contain sensors to help them interact with the environment around them, and the data they collect will create many opportunities.
For instance, this will allow the improvement of strategies for resource usage and management in settings such as urban planning and environmental sustainability, thus advancing agriculture and increasing the number of smart cities.
It will also promote automation and the use of cyber-physical systems, prompting the widespread adoption of industry 4.0 practices.
Moreover, considering that important processes such as data analytics can benefit from working with more data, this will lead to data scientists not only being able to better understand the world we live in, but also making more accurate predictions and creating improved systems based on people's behaviors and tastes.

When working with the data streams generated by sensors, it is common for users to need to transform the raw collected data into something else before storing the final values.
For example, they may wish to standardize or clean the data (parsing), discard unwanted readings (filtering), or generate knowledge (predicting).
Currently, the cloud is the usual place of choice for executing this type of task.
However, with the ever-growing scale of the IoT, the global network infrastructure will face the challenge of handling a rise in demand that may reach the order of petabytes every day.
In this scenario, approaches such as sending all data to be processed and stored in the cloud will become too taxing in terms of communication time, financial cost, and energy consumption.

A possible way to meet these expected requirements is not sending all the data to be processed by machines that are far from the data source, but instead bringing the computation closer to where the information already is.
With this premise, a new paradigm called fog computing emerged, proposing to process data closer to network edge devices, such as switches and routers~\cite{Bonomi2012-Fog_Role_IOT}.

As the IoT becomes more widespread, a large growth in the number and importance of IoT devices with limited resources (e.g., power, memory, or processing) is expected. 
One evidence that points toward this is the fact that the USA's National Institute of Standards and Technology (NIST) has already called attention to this type of device within the fog computing context, naming them mist nodes~\cite{NIST2018-FogComputingConceptualModel}, as they are lightweight fog computing nodes.
Therefore, to reach the full capabilities of the IoT, both industry and academia must explore the full potential of these devices.

However, we point out one main difference between our approach and the characterization made by NIST.
While they define mist nodes as more specialized, dedicated, and often sharing the same locality with the smart end-devices they service, we consider a scenario where constrained devices are integrated with sensors and are capable of general-purpose code execution, thus being able to run custom code sent by users.
This way, mist nodes would not be simply nearby devices that can perform specific tasks to process sensor data, but, instead, be devices that can be leveraged by users interested in analyzing the data they collect.

The main contributions of this chapter are as follows:

\begin{itemize}
	\item A literature review on the categorization of constrained devices, placing them in the context of fog computing and the IoT.
	\item A classification of real-world devices according to the constrained device's categorization presented.
	\item An overview of the current IoT and fog computing scenarios.
	\item A categorization of applications relative to their compatibility with constrained devices.
	\item A discussion about future possibilities for the IoT considering the use of fog computing and constrained devices.
	\item A description of the challenges and opportunities brought by the topics explored in this chapter.
\end{itemize}

\section{Background}
This section outlines the key ideas to which this chapter is related.
The aim here is not to cover each topic exhaustively, but rather lay out important concepts that provide a broader view of the discussion presented in this chapter.

\subsection{Constrained Devices Classification}
\label{subsec:constrained_devices}
The concept of a \emph{constrained device}\index{constrained device}, or \emph{resource-constrained device}\index{resource-constrained device}, relates to restrictions imposed on power, energy, memory, and processing resources of small devices.
As there is still no well-established threshold for the size of these constraints, we chose to adopt the definition presented in the ``Terminology for Constrained-Node Networks'' document~\cite{Bormann2014-TerminologyConstrained}, which describes several terms with the intent to help the standardization work for constrained-node networks and represents the consensus of the Internet Engineering Task Force (IETF) community.

According to the report, a constrained device is one that does not have some of the characteristics that are expected to be present on other devices currently connected to the Internet.
This is often due to cost constraints and/or physical restrictions of features such as size, weight, and available power and energy.
The tight limits on power, memory, and processing resources lead to hard upper bounds on power state, code space, and processing cycles, making the optimization of energy and network bandwidth usage a primary concern in all design requirements.
Although this may not be a rigorous definition, the IETF created it based on the state of the art and it can clearly separate constrained devices from server systems, desktop and laptop computers, and more powerful mobile devices such as smartphones~\cite{Bormann2014-TerminologyConstrained}.

The IETF proposed classifications for constrained devices considering the size of available memory, energy limitations, and strategies for power usage and network communication.
Below is the proposed categorization with visual aids to better convey the expected behavior of the devices that belong to the classes in each category.

\begin{figure*}[htpb]
	\centering
	\includegraphics[width=0.85\textwidth]{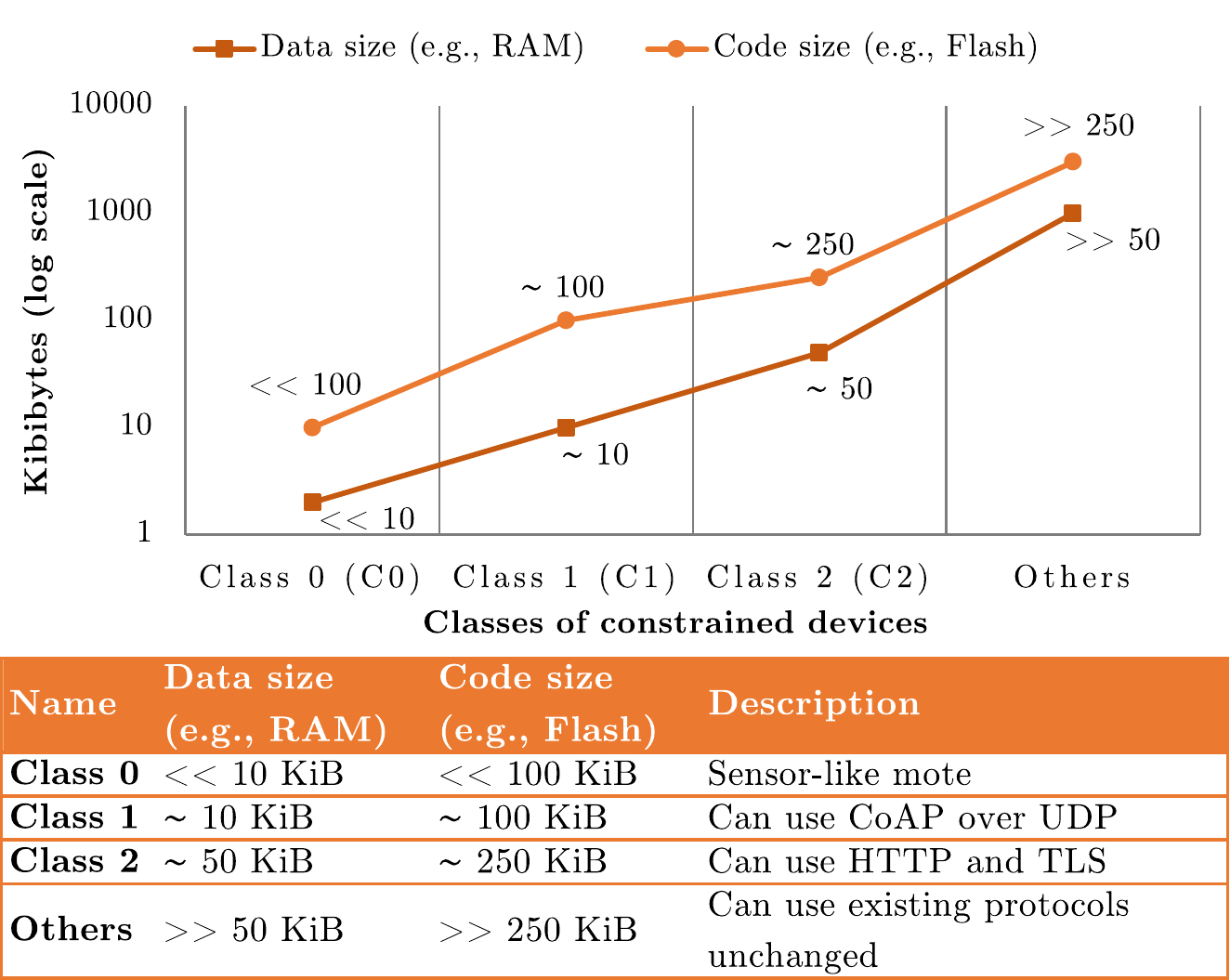}
	\caption{Approximate boundaries for the classes of constrained devices (lower = more constrained).}
	\label{fig:classes_C}
\end{figure*}

Figure~\ref{fig:classes_C} shows the classification for constrained devices.
It is important to note that, while the boundaries presented in this classification\footnote{The constrained devices classification was defined using the units established by the International Electrotechnical Commission~(IEC), where 1~kibibyte~(\si{\kibi\byte}) = 1024 bytes and 1~mebibyte~(\si{\mebi\byte}) = 1024 kibibytes.} are expected to change over time, Moore's law tends to be less accurate when it comes to embedded devices in comparison to personal computing settings.
This can be explained by the fact that gains obtained by increasing transistor count and density in embedded devices are more likely to be invested in reducing cost and power requirements than into continual increases in computing power.

\textit{Class 0 (C0)} contains sensor-like devices.
Although they may answer keep-alive signals and send on/off or basic health indications, they most likely do not have the resources to securely communicate with the Internet directly (larger devices act as proxies, gateways, or servers) and cannot be secured or managed comprehensively in the traditional sense.

Devices that belong to \textit{Class 1 (C1)} are quite constrained in code space and processing capabilities and cannot easily talk to other Internet nodes employing a full protocol stack (e.g., HTTP, TLS).
Nonetheless, they can use the Constrained Application Protocol (CoAP) over UDP, participate in meaningful conversations without the help of a gateway node, and be integrated as fully developed peers into an IP network.

\textit{Class 2 (C2)} devices are less constrained and capable of supporting most of the same protocol stacks as servers and laptop computers.
They can benefit from lightweight and energy-efficient protocols and from consuming less bandwidth.
In addition, using the protocol stacks defined for more constrained devices may reduce development costs and increase the interoperability of these devices.

Devices with capabilities significantly beyond that of Class 2 are left uncategorized (\textit{Others}).
They may still be constrained by a limited energy supply, but can largely use existing protocols unchanged.

The graph in Figure~\ref{fig:classes_C} shows the approximate order of magnitude of data~(e.g., RAM) and code~(e.g., Flash) memory sizes of the devices in each of these classes.
It is possible to notice that there is a significant difference in resources between C0 and uncategorized devices.
Moreover, it is important to point out that this classification creates a relevant distinction from one class to another in practice, as memory size has an impact on the communication protocols that the device can use.

\begin{figure*}[htpb]
	\centering
	\includegraphics[width=0.85\textwidth]{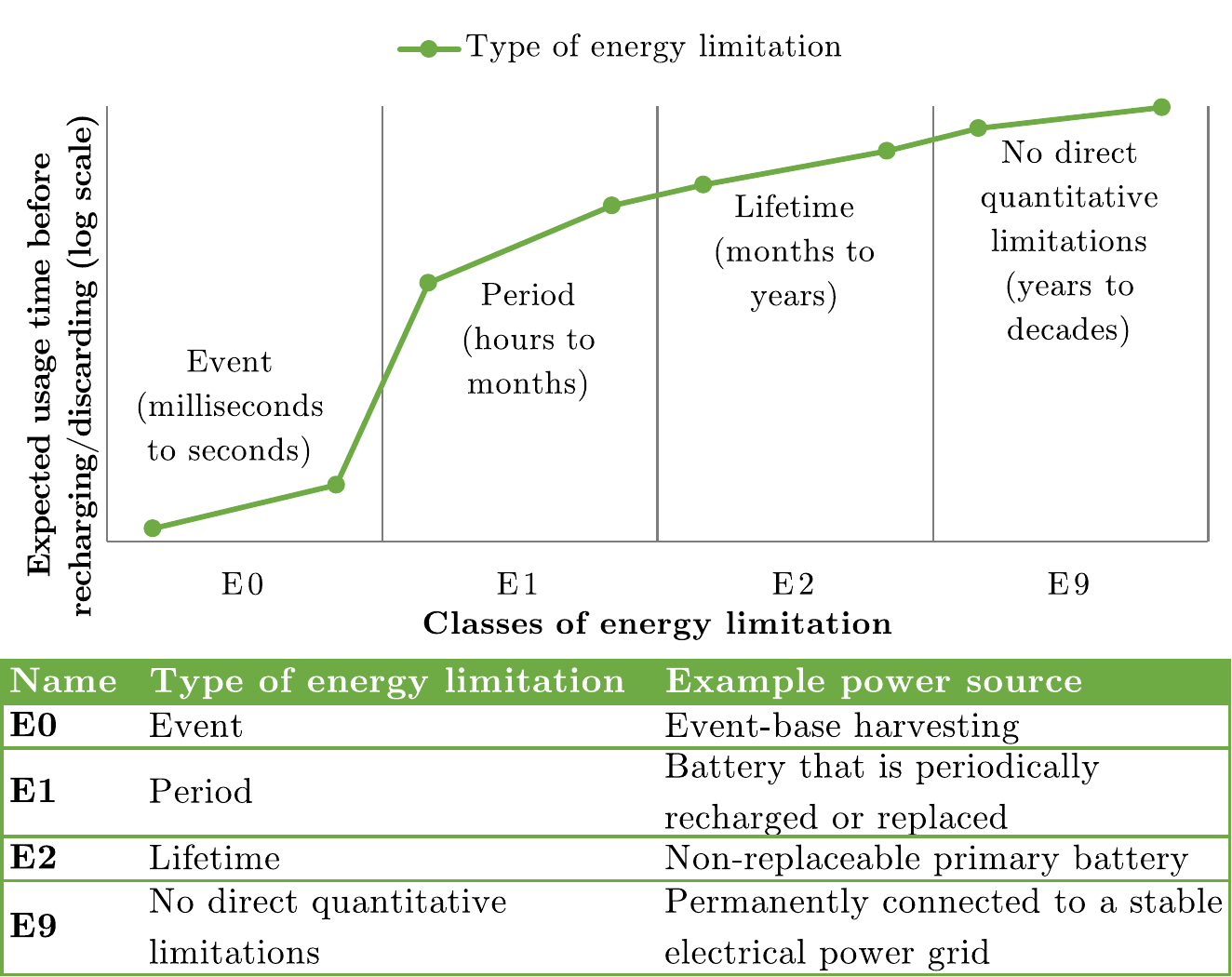}
	\caption{Expected behavior of classes of energy limitation in terms of time before recharging/discarding (lower = more constrained).}
	\label{fig:classes_E}
\end{figure*}

The second classification, which Figure~\ref{fig:classes_E} presents, pertains to limitations to the total electrical energy available before the device's energy source is exhausted.

Devices that belong to \textit{Class E0} depend on event-based harvesting and have a limited amount of energy available for a specific event (e.g., a button press in an energy-harvesting light switch).

\textit{Class E1} devices have batteries that can be periodically recharged or replaced, which leads to relevant limitations on specific periods (e.g., a solar-powered device with a limited amount of energy available for the night, a device that is manually connected to a charger and has a period between recharges, or a device with a periodic primary battery replacement interval).

Devices from \textit{Class E2} rely on non-replaceable primary batteries, thus having their total energy available limited by the usable lifetime of the device (e.g., a device that is discarded when its non-replaceable primary battery is exhausted).
In a sense, many E1 devices are also E2, as the rechargeable battery has a limited number of useful recharging cycles.

Devices that are permanently connected to a stable electrical power grid have no relevant limitations concerning energy and thus are placed on \textit{Class E9}.

The graph in Figure~\ref{fig:classes_E} has the expected order of magnitude for the period in which the device will work connected to the power source that describes its class before needing to be recharged or discarded.
In the case of events, is expected the interaction to last from milliseconds to seconds.
As for periods, the device may work for several hours up to a few months before it needs to be recharged.
It is possible to anticipate that lifetime devices will perform from months to years before needing replacement, and devices connected to a stable power grid should work for many years or even a few decades before failing for possibly unrelated reasons.
In practice, it is important to differentiate between these classes because the type of energy source used by the device affects the types of applications for which this device can be used.

\begin{figure*}[htpb]
	\centering
	\includegraphics[width=0.85\textwidth]{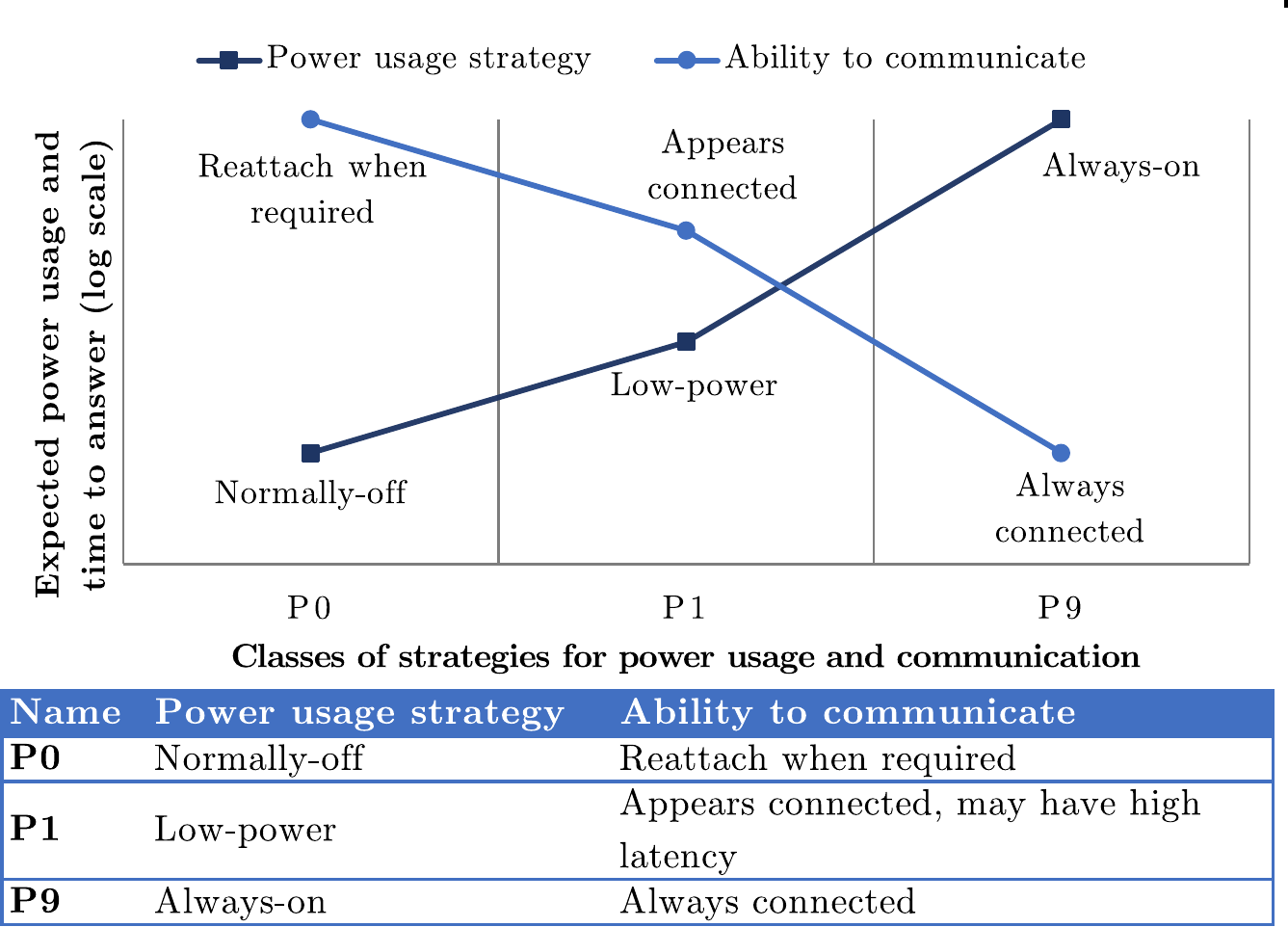}
	\caption{Expected behavior of classes of strategies for power usage in terms of power usage (lower = more constrained) and communication in terms of time to answer (higher = more constrained).}
	\label{fig:classes_P}
\end{figure*}

Lastly, it is also possible to classify constrained devices according to the strategies they employ for power usage and network attachment.
Figure~\ref{fig:classes_P} displays this classification.

\textit{Class P0} devices sleep for long periods and reattach to the network as they are woken up (normally-off).
If communication is infrequent, the relative increase in energy expenditure during reattachment may be acceptable.
The main optimization goal, in this case, is to minimize the effort during the reattachment process and any resulting application communications.

The strategy of \textit{Class P1} (low-power) is most applicable to devices that operate on a very small amount of power but still need to be able to communicate rather frequently.
The device may retain some form of attachment to the network due to the small extent of time between transmissions.
In this case, it is possible to optimize parameters such as the frequency of communication and perform \emph{duty cycling} (i.e., regularly switching components on and off).

The approach used for \textit{Class P9} (always-on) is most applicable if extreme power-saving measures are not required.
The device can always stay on as usual for non-constrained devices and be connected to the network.
Power-friendly hardware or limiting the number of wireless transmissions, CPU speed, and other aspects may be useful for general power-saving and cooling needs.

The graph in Figure~\ref{fig:classes_P} shows two different aspects of the expected behavior of these devices.
First, we have the expected power usage, which grows along with the time that the device stays on.
It is important to note that this is due to the communication frequency that is being assumed for each of these classes, as using a normally-off strategy in a situation where communication is very frequent may cause the device to spend a lot of power by executing continual shutting down and powering on processes.
Then, we have the time necessary for the device to answer requests, which is larger in cases where more aggressive power-saving techniques are used.
We expected this delay to be much more prominent on devices that need to be woken up for communication, with low-power devices also having high latency.
Devices that are always connected should not present any meaningful lag unless network availability is disrupted.
The placement of the points in this graph does not intend to show specific quantities, but rather what we view as the general tendency in the behavior of these devices.
Like the two previous classifications, the measures used by this categorization also have an important real-world impact when deciding which devices are suitable for certain applications.

\begin{figure*}[htpb]
	\centering
	\includegraphics[width=\textwidth]{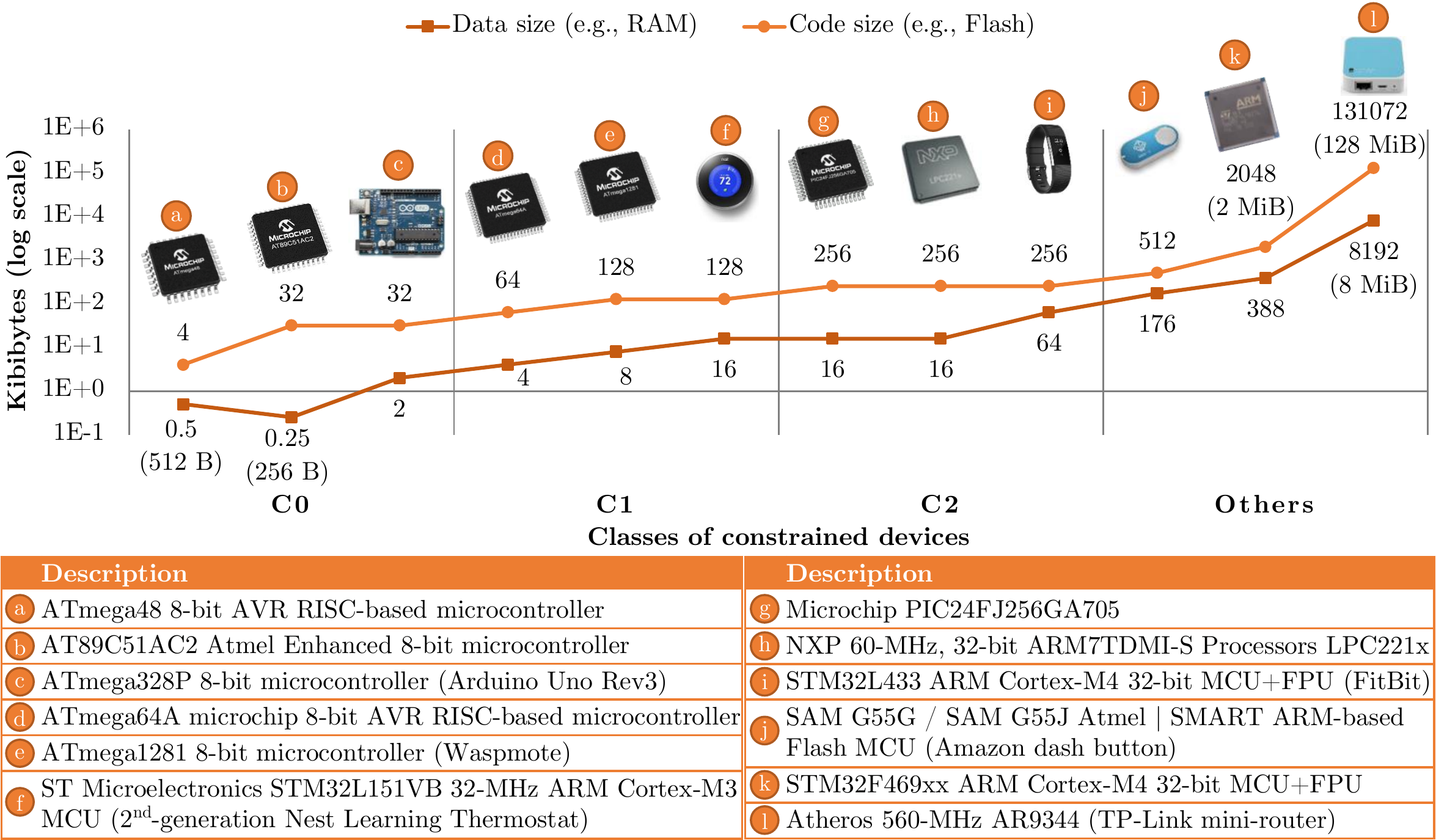}
	\caption{Categorization of real-world devices according to the constrained device's classification.}
	\label{fig:device_classification}
\end{figure*}

Furthermore, we can think of real-world devices and how they fit into the constrained devices classification.
Figure~\ref{fig:device_classification} shows twelve current devices, which are either uncategorized or belong to one of the classes C0, C1, or C2.
In particular, we highlight the popular devices that belong to each category: in C0, we have the Arduino Uno board, which is the most-used board in the Arduino family; the \nth{2}-generation Nest Learning Thermostat belongs to C1 and is a smart device that programs itself automatically and helps homeowners save energy; and, in C2, we have the Fitbit activity tracker, which assists people with exercising, eating, and sleeping habits.

\subsection{Internet of Things}
\label{subsec:iot}
When the \textit{Internet of Things}\index{Internet of Things} was first proposed, the focus was to create a network connecting objects that would be interoperable and uniquely identifiable by Radio-Frequency Identification (RFID).
Since then, this definition has evolved to include ideas that work on a much larger scope: the IoT must be a dynamic network infrastructure with self-configuring capabilities based on standards.
In addition, its components are now seen as physical or virtual ``things'' that have attributes as well as identities and are able to use intelligent interfaces and be part of an information network~\cite{Li2015-IoTSurvey}.

Nevertheless, the IoT is still in its infancy, thus opening it up to different definitions.
For instance, we may also expect the components of the IoT to have sensing and processing capabilities, allowing everyday objects to not only communicate with one another but also work with services available on the Internet to achieve useful goals~\cite{Whitmore2015-IoTSurvey}.

The ubiquity of objects such as mobile phones, sensors, actuators, and RFID tags will create a scenario where the majority of content will be created and received by these ``things'', surpassing the scale that we currently see on the Internet, and leading to a system that ensures connectivity anytime, anywhere, for anyone and anything~\cite{Atzori2010-IoTSurvey}.

The key to the innovation of the IoT lies precisely in its never-seen-before size.
While there already exist embedded systems connected to the Internet in modern-day appliances, the IoT presents the perspective of billions or trillions of smart objects that bring about the creation of a smart planet where people can benefit from the intelligence gathered by combining information coming from different sources~\cite{Kopetz2011-IoT}.

There are several areas in today's society that can profit from the wide deployment of the IoT.
For didactic purposes, it is possible to group them into different domains: healthcare, novelty, personal and social, smart environments, smart infrastructure, and transportation and logistics.

\emph{Healthcare}\index{Internet of Things!healthcare} applications for the IoT propose to improve the quality of life by allowing medical professionals to provide better service with reduced costs~\cite{zeadally2019IoT}.
Automatic \emph{data collection} from patients and hospital property can reduce form-processing time, as well as facilitate procedure auditing and inventory management.
Patient \emph{identification and authentication} can prevent incidents such as the incorrect administration of drugs and dosages and procedures being executed on the wrong patient.
It can also ease the identification of newborn children in maternities.
Using the IoT to identify hospital assets can avoid important instruments being misplaced or stolen, and staff authentication can improve safety inside the premises~\cite{Atzori2010-IoTSurvey}.
In fact, \emph{medication} delivery can even be automated by smart implants~\cite{Kopetz2011-IoT}, which can also monitor allergic reactions and adverse interactions~\cite{Whitmore2015-IoTSurvey}.
Continuously \emph{monitoring} the patient's vital statuses (e.g., heart rate, blood pressure, body temperature, glucose level) and combining this information to form a comprehensive picture of their health at each moment can improve treatment and responsiveness, even if the patient is not in the hospital~\cite{Kopetz2011-IoT,Whitmore2015-IoTSurvey,Li2015-IoTSurvey,Atzori2010-IoTSurvey}.
Body area networks can \emph{track} daily activities for patients in assisted living and trigger message alarms in case of emergencies~\cite{Li2015-IoTSurvey,Kopetz2011-IoT}.
Tracking the position of patients can also improve the workflow inside the hospital and tracking materials can prevent items from being left inside the patient during surgery~\cite{Atzori2010-IoTSurvey}.

\emph{Novelty}\index{Internet of Things!novelty} IoT applications are those that depend on technologies that are currently under development but are expected to not be fully adopted in the short/medium term.
Examples of this type of application are: \emph{augmented reality}, where smartphones or other devices can provide background information about smart objects by accessing several context-dependent databases~\cite{Kopetz2011-IoT}; \emph{city information models}, which monitor the status of buildings and structures such as sidewalks, sewers, and railways, providing information about environmental performance and enabling them to share resources like energy in the most efficient and cost-effective way; \emph{enhanced game rooms}, which measure the excitement and energy of the players by sensing movement, noise, heart rate, blood pressure, etc., and then set achievements and adjust the game accordingly; and \emph{robot taxis}, which operate with or without a driver, providing service in a timely manner, responding to real-time traffic updates in the city, and automatically returning to set points for maintenance, recharging, and cleaning~\cite{Atzori2010-IoTSurvey}.

There are many applications for the IoT in \emph{personal and social}\index{Internet of Things!personal}\index{Internet of Things!social} settings.
They enable users to interact with other people, maintain and build social relationships, and manage their personal needs.
For example, IoT-enabled mobile phones could connect to each other and \emph{share contact information} when their owners have matching dating or friendship profiles~\cite{Whitmore2015-IoTSurvey}.
\emph{Missing objects} could easily be located using RFID tags and a search engine that queries their last recorded position.
User-defined events can also notify owners whenever an object's location matches some predefined conditions.
A similar use case for this technology would be \emph{preventing thefts}, as the stolen items would be able to inform their owner that they are being relocated without authorization.
Furthermore, it would be feasible for sensors to record events that happen to people or objects and build timelines that enable \emph{historical queries}.
Users would be able to look back and see how and with whom they have spent their time and trend plots could be automatically generated to help them organize their future schedule~\cite{Atzori2010-IoTSurvey}.
Another possible application would be tracking user \emph{location} and letting them know when they are close to friends, social events, or other interesting activities~\cite{Whitmore2015-IoTSurvey}.
Finally, IoT devices can be integrated with \emph{social networks} to help users save time by providing automatic real-time information about their activities and locations~\cite{Atzori2010-IoTSurvey,Whitmore2015-IoTSurvey}.

The intelligence of objects in the IoT can help bring more comfort to people through the creation of \emph{smart environments}\index{Internet of Things!smart environment}.
For instance, \emph{homes and offices} can become more \emph{comfortable} by having heating that adjusts itself to the weather and lighting regulated by the time of the day~\cite{Atzori2010-IoTSurvey, filho2019IoT}.
It is also possible to \emph{reduce wastes and costs} by tracking the usage of objects and resources like electrical energy and then use this data to make decisions such as turning off pieces of equipment when they are unneeded~\cite{Atzori2010-IoTSurvey,Li2015-IoTSurvey,Whitmore2015-IoTSurvey}.
The IoT can enhance other spaces as well.
Mass deployment of RFID tags can enable automation in \emph{industrial plants}~\cite{Atzori2010-IoTSurvey} and reduce maintenance costs and safety issues by detecting equipment anomalies before they lead to failures~\cite{Kopetz2011-IoT}.
\emph{Gyms and museums} are good examples of smart leisure environments, as the IoT can help these facilities to provide a more customized service.
In the gym, exercise machines can change their parameters for each person and monitor their health during the training session, while the museum can adjust the climate of each of its rooms corresponding to their current exhibition~\cite{Atzori2010-IoTSurvey}.

We can also have a \emph{smart city infrastructure}\index{Internet of Things!smart city infrastructure}.
By spreading sensors around public areas, it is possible to monitor many factors that affect life in the city.
To name a few, there is the \emph{air quality}, \emph{traffic}, \emph{availability of parking spaces}, and whether \emph{trash containers} are full or not.
This information enables the development of more efficient city services, leading to a better quality of life for its inhabitants.
Integrating smart objects into the \emph{urban physical infrastructure} also contributes to improving flexibility, reliability, and efficiency in infrastructure operation, as well as increasing safety and reducing both the costs and the number of workers necessary for building and maintenance~\cite{Whitmore2015-IoTSurvey}.
Another application is IoT-based \emph{surveillance}, which can enhance overall security for the population.
Lastly, \emph{smart grids} can help to better coordinate energy supply and demand with smart meters, as detecting energy usage patterns in the collected data allows the creation of plans to reduce energy consumption~\cite{Kopetz2011-IoT,Whitmore2015-IoTSurvey, hossain2019IoT}.

\emph{Transportation and logistics}\index{Internet of Things!transportation}\index{Internet of Things!logistics} can both benefit greatly from the IoT.
If public and private means of transportation were equipped with sensors, actuators, and processing power, and perhaps even the roads themselves with tags and sensors, it would be possible to see considerable advances in technologies such as \emph{assisted driving}~\cite{Olakanmi2019IoT}.
Route optimization would be improved through automatic real-time reports about traffic jams and incidents, and systems for collision avoidance and accurate monitoring of important vehicles (e.g., the ones transporting hazardous materials) could be implemented.
Overall, providing drivers with more information about traffic and road conditions would lead to better navigation and safety.
Near-Field Communication (NFC) tags and visual markers in posters and panels could facilitate \emph{business transactions} such as purchasing tickets for public transportation, since hovering a mobile phone over a tag or pointing it toward a marker would allow users to find information about services such as costs and available seats and buy tickets.
NFC could also \emph{augment touristic maps}, centralizing up-to-date information about hotels, restaurants, monuments, and events in the tourist's area of interest~\cite{Atzori2010-IoTSurvey}.
Supply chains have already made some efforts to increase the efficiency of \emph{logistics} with technologies like RFID.
However, the wide deployment of IoT-connected, low-cost RFID tags and readers in packages, delivery trucks, and storehouses would enable the sharing of real-time information with consumers and across companies, regardless of geographical distance~\cite{Whitmore2015-IoTSurvey}.
Improved traceability of products would result in better information about the availability of items at stores and more precise delivery time estimates~\cite{Atzori2010-IoTSurvey,Whitmore2015-IoTSurvey}.
It would also enhance the management of shelf space and inventory, enabling automatic stocking according to demand~\cite{Whitmore2015-IoTSurvey,Kopetz2011-IoT,Atzori2010-IoTSurvey}.
Reducing human involvement in the storage and delivery process could contribute to shorter completion times for sales as well~\cite{Kopetz2011-IoT,Atzori2010-IoTSurvey}.
Another effect of having items that can be traceable at all times is \emph{preventing counterfeiting} of, for example, products and currency bills~\cite{Whitmore2015-IoTSurvey}.
In addition to tracking items, supply chains can also \emph{monitor the environment} of perishable foods considering parameters like temperature, humidity, and shock in order to assure quality levels from production to consumption~\cite{Atzori2010-IoTSurvey}.

\begin{figure}[htpb]
	\centering
	\includegraphics[width=\textwidth]{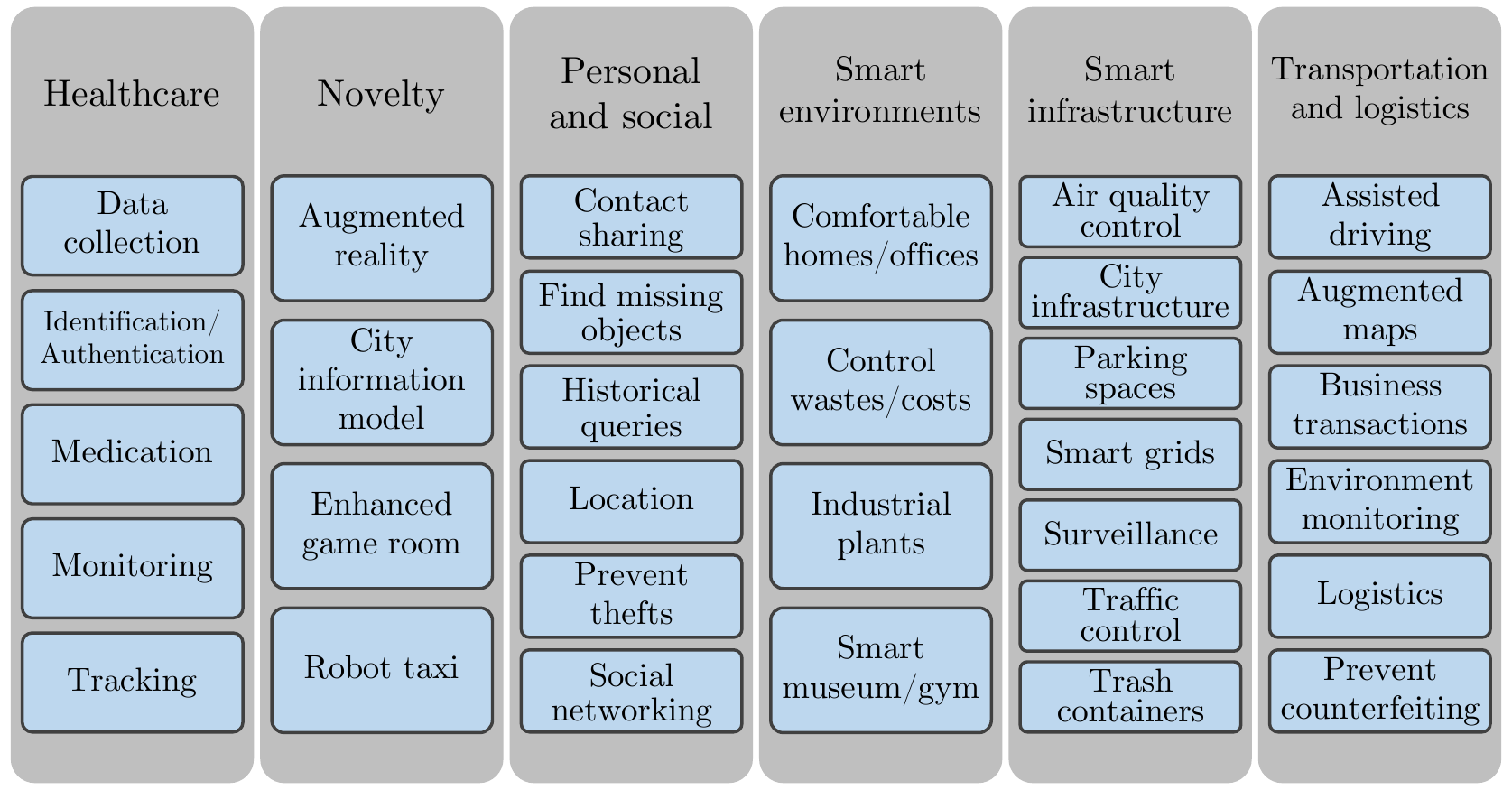}
	\caption{Applications for the Internet of Things.}
	\label{fig:iot_applications}
\end{figure}

Figure~\ref{fig:iot_applications} summarizes the IoT applications presented in this subsection.
However, it is important to emphasize the fact that many more uses for this technology will emerge as its development matures.

Lastly, it is important to note that the Internet of Things will work as a complement to the cloud, not a replacement.
Smart objects will be able to use the services provided by the cloud and the tasks attributed to each of them will depend on several factors.
For instance, there is privacy, since sending sensitive data to third-party servers risks the information not being used as agreed; energy consumption, as executing a task locally may or may not require more energy than sending its parameters to a data center; autonomy of the smart objects, seeing that it would be unfeasible to work with a huge number of objects that require frequent human intervention; response time, given that some latency requirements will not be met when the time necessary for sending data to and receiving data from the cloud is considered; reliability, for tasks are more prone to failure if they are distributed to a wide range of machines that are not very robust; and security, considering that sending data to the cloud opens it up to attacks, possibly jeopardizing confidentiality and integrity~\cite{Kopetz2011-IoT}.

\subsection{Fog Computing}
\label{subsec:fog}
As in many other emerging technologies, fog computing\index{fog computing} is a concept that is still under discussion~\cite{Chiang2017-10questions}.
Concisely, we can define it as a paradigm that aims at filling a gap between the cloud and end-devices.
However, there is a wide range of terms and definitions in the literature that address similar approaches.

In 2012, Bonomi et al.~\cite{Bonomi2012-Fog_Role_IOT} from Cisco Systems defined fog computing as a ``highly virtualized platform that provides compute, storage, and networking services between end devices and traditional cloud Computing Data Centers, typically, but not exclusively located at the edge of network''.

The authors claim that the origins of the fog can be traced to early proposals to support endpoints with rich services at the edge of the network, including applications with low latency requirements, such as gaming, video streaming, and augmented reality~\cite{BITTENCOURT2018, Gama2018, Immich2018}.
As fog and cloud computing share many similarities, they credit the creation of the term to the definition of fog as a cloud that is close to the ground.
In the same text, they also present a comprehensive list of defining characteristics for the fog: low latency and location awareness, widespread geographical distribution, mobility~\cite{Bittencourt2017-MobilityFog}, a very large number of nodes, the predominant role of wireless access, the strong presence of streaming and real-time applications, and heterogeneity.

Additionally, Vaquero and Rodero-Merino~\cite{Vaquero2014-FindingYourWayInTheFog} pointed out in 2014 that the novelty of the fog is a result of many new trends in technology usage patterns (e.g., the need for network connectivity and privacy for an extremely large number of small devices) and the advance of enabling technologies (e.g., 3D microbatteries, new ways to manage networks through software, and new wireless communication protocols).
They added ubiquity, improved network capabilities as a hosting environment, and better support for cooperation among devices to the list of important features for fog computing.

Skala et al.~\cite{Skala2015-Scalable_Hierarchy_Cloud_Fog_Dew} proposed an even further separation of the distributed computing hierarchy in 2015, placing a new structural layer called dew computing below the fog with the intent of scattering information among end-user devices so data can be accessed even when an Internet connection is unavailable.

In November of 2015, the OpenFog Consortium was founded with the goal of establishing an open architectural framework that can help business leaders, software developers, hardware architects, and system designers to create and maintain the hardware, software, and system elements necessary for implementing fog computing~\cite{OpenFog2017-OpenFogReferenceArchitecture}.
Since then, they have published many reports such as a white paper~\cite{OpenFog2016-OpenFogArchitectureOverview} and a Reference Architecture for fog computing~\cite{OpenFog2017-OpenFogReferenceArchitecture} to foster the creation of fully interoperable and secure systems.
At the beginning of 2019, the OpenFog Consortium and the Industrial Internet Consortium decided to work together under the Industrial Internet Consortium name~\cite{OpenFog-IndustrialInternet}.

Other entities, such as the USA's NIST, are also making efforts to help define the fog and, to this end, they published a Conceptual Model~\cite{NIST2018-FogComputingConceptualModel} that places fog computing in relation to cloud and edge computing.
They characterize fog computing as ``a layered model for enabling ubiquitous access to a shared continuum of scalable computing resources,'' saying that it ``facilitates the deployment of distributed, latency-aware applications and services, and consists of \textit{fog nodes} (physical or virtual), residing between \textit{smart} end-devices and centralized (cloud) services.''
These fog nodes ``can be organized in clusters - either vertically (to support isolation), horizontally (to support federation), or relative to \textit{fog nodes}' latency-distance to the \textit{smart} end-devices.''~\cite{NIST2018-FogComputingConceptualModel}.
In this chapter, we adopt this definition of fog computing.

It is important to emphasize that, much like the IoT, the fog does not intend to supersede the cloud but rather complement it~\cite{Bonomi2014-Fog_Platform_IoT_Analytics}.
The cloud has shortcomings that could be mitigated by the fog such as requiring high bandwidth client access links and being unable to meet the latency and security requirements of certain problems~\cite{Firdhous2014-Fog_Future_Cloud}.
Due to its characteristics, the fog is a good approach to handling large-scale distributed control systems and geo-distributed or time-sensitive mobile applications~\cite{Bonomi2014-Fog_Platform_IoT_Analytics}.

There are many situations where it is possible to use the fog and the cloud in conjunction, leveraging both fog localization and cloud globalization.
For instance, we can consider the IoT applications presented in Section~\ref{subsec:iot}.
The majority of the use cases described can benefit from the fog's low latency, while cases such as data collection, historical queries, control wastes/costs, and city infrastructure can use it to filter and pre-process streaming data.
This combination is particularly favorable, or perhaps even required, for analytics and Big Data~\cite{Bonomi2012-Fog_Role_IOT}, as the IoT has brought a new magnitude to these fields due to its enormous number of devices distributed at the edge of the network~\cite{Bonomi2014-Fog_Platform_IoT_Analytics}.

Moreover, the hierarchical structure envisioned for the fog can provide many advantages when we think about its increased potential for scalability and its ability to reduce response latency.
By enabling infrastructures to continue working even when the connection to the cloud is unavailable, the fog increases their reliability and availability.
As each fog layer addresses data security and privacy, the fog allows for solutions that are overall more secure as well (e.g., it is possible to aggregate privacy-sensitive data before they even leave the network edge)~\cite{Byers2017-FogArch}.

We can also see fog nodes as part of a compute fabric that provides high-performance computing and massive processing.
Currently, this fabric is mainly formed by cloud computing services, but fog nodes are capable of enhancing its functionalities, as they are cheaper and offer lower latency~\cite{Bittencourt2017-MobilityFog}.
Furthermore, when we consider the presence of resource-limited IoT devices, it is possible to notice that this compute fabric may start at large servers on the cloud and extend all the way through the fog hierarchy toward the devices that are collecting the data.
Despite the heterogeneity within the fabric in this scenario, the expectation is that applications will execute in different devices seamlessly.

In order to better place the fog in the current distributed computing scenario, we compare the features of this paradigm with other approaches: the cloud, cloudlets, edge computing, Mobile Cloud Computing (MCC), Multi-access Edge Computing (MEC), Wireless Sensor Networks (WSNs), and Wireless Sensor and Actuator Networks (WSANs).

As mentioned before, it is easy to see that the cloud\index{cloud computing} and the fog share many similarities, with the major difference being that the latter is closer to end-users.
However, a more in-depth look reveals that this divergence can affect a series of characteristics in both technologies.
Firdhous, Ghazali, and Hassan~\cite{Firdhous2014-Fog_Future_Cloud} made such a comparison and Table~\ref{tab:fog_cloud_comparison} presents a summary of their results.
From this analysis, we again see that fog computing excels in time-sensitive tasks while the cloud still presents advantages for more demanding jobs such as processing batches.

\begin{table}[htpb]
	\caption{Comparison between cloud and fog computing. Modified from~\cite{Firdhous2014-Fog_Future_Cloud}.}
	\label{tab:fog_cloud_comparison}
	\centering
	\begin{tabular}{L{4cm}L{3cm}L{3cm}}
		\toprule
		Characteristic            & Cloud computing     & Fog computing \\
		\toprule
		Latency                   & High                & Low \\
		\hline
		Delay jitter              & High                & Very low \\
		\hline
		Location of server nodes  & Within the Internet & Edge of the LAN \\
		\hline
		Client-server distance    & Multiple hops       & One hop \\
		\hline
		Security                  & Undefined           & Can be defined \\
		\hline
		Attack on data en-route   & High probability    & Very low probability \\
		\hline
		Location awareness        & No                  & Yes \\
		\hline
		Geographical distribution & Centralized         & Distributed \\
		\hline
		Number of server nodes    & Few                 & Very large \\
		\hline
		Computational power of
		server nodes    & High  & Low \\
		\hline
		Support for mobility      & Limited             & Supported \\
		\hline
		Real-time interactions    & Supported           & Supported \\
		\hline
		Last mile connectivity    & Leased line         & Wireless \\
		\bottomrule
	\end{tabular}
\end{table}

Stojmenovic~\cite{Stojmenovic2014-FogCloudToGround} defines \emph{cloudlets}\index{cloudlet} as layers placed between the cloud and mobile devices that operate as ``data centers in a box'', that is, they are often owned by a local business and have little or no professional attention.
When working with these intermediate layers, mobile devices play the role of thin clients and are able to rapidly instantiate custom virtual machines to perform the required computation.
This way, they avoid part of the overhead brought by the use of the cloud and therefore minimize their response time.

Akin to the fog, the key behind this idea is to bring the cloud closer to the data, but, in the case of the cloudlets, this only affects a few users at a time.
Both models are also similar in terms of the demand for new solutions for issues such as privacy, software licensing, and business models.

Although cloudlets were introduced in 2009~\cite{Satyanarayanan2009-Cloudlet}, years before Bonomi et al.~\cite{Bonomi2012-Fog_Role_IOT} presented their definition of fog computing, an analysis of the characteristics of both platforms points to cloudlets being an important special case of fog computing.

The concept of \emph{edge computing}\index{edge computing} is also very similar to the fog and sometimes they are even used interchangeably~\cite{Yi2015-SurveyFog}.
Nevertheless, edge computing was created with the focus of distributing Web applications on the Internet's edge~\cite{Davis2004-Edge}, so it is considered that it represents another subset of fog computing.

\emph{MCC}\index{Mobile Cloud Computing} is defined as an infrastructure where the data are both stored and processed outside of mobile devices.
It relies on the cloud to provide processing power and data storage, thus allowing smartphone users and other mobile subscribers to benefit from mobile cloud applications and mobile computing~\cite{Yi2015-SurveyFog}.

\emph{MEC}\index{Multi-access Edge Computing}, formerly known as Mobile-Edge Computing, is a natural development in the evolution of mobile base stations and related communication technologies to support a cloud server running at the edge of the network. 
Using local content and real-time information about Local Area Network (LAN) and Radio Access Network~(RAN) conditions, this technology enables the execution of tasks that cannot be handled by traditional infrastructures due to issues like network congestion~\cite{Yi2015-SurveyFog,Taleb2017}.

While fog computing may seem like a combination of MCC and MEC, the fact that it is applied to a broader set of devices and networks makes it stand out as a more promising and well-generalized computing paradigm in the context of the IoT~\cite{Yi2015-SurveyFog}.

\emph{WSNs}\index{Wireless Sensor Network} usually connect a large number of low bandwidth, low energy, low processing power, small memory motes, i.e., wireless sensor nodes that run on extremely low power or even extend their battery life by harvesting energy.
Each mote on a WSN acts as sources to one or more sinks (collectors).
This type of network is suitable for tasks such as sensing the environment, performing simple processing, and sending data to the sink.
However, WSNs fall short in cases that not only require sensing and tracking, but also performing physical actions with the help of actuators (e.g., opening, closing, targeting, or even carrying and deploying sensors).

Actuators, which can control either a system or the measurement process itself, bring a new dimension to sensor networks, as the communication of \emph{WSANs}\index{Wireless Sensor and Actuator Network} goes from both sensors to sinks and from controller nodes to actuators.
Unlike networks that only connect motes, issues of stability and possible fluctuations in behavior cannot be ignored in this case, and latency and jitter become the main concerns if the system also requires that the actions be performed in near real time.

If we consider that fog computing is characterized by proximity and location awareness, geo-distribution, and hierarchical organization, we see that it is a suitable platform to support energy-constrained WSNs and WSANs~\cite{Bonomi2012-Fog_Role_IOT}.

Overall, the comparison between fog computing and other current technologies leads us to the conclusion that, in each case, the fog either represents a concept that is broader than previous ideas or a complementary platform that enables previous approaches to thrive within the immense scale of the Internet of Things.

\subsection{Mist Computing}
\label{subsec:mist}
\begin{figure*}[htpb]
	\centering
	\includegraphics[width=0.95\textwidth]{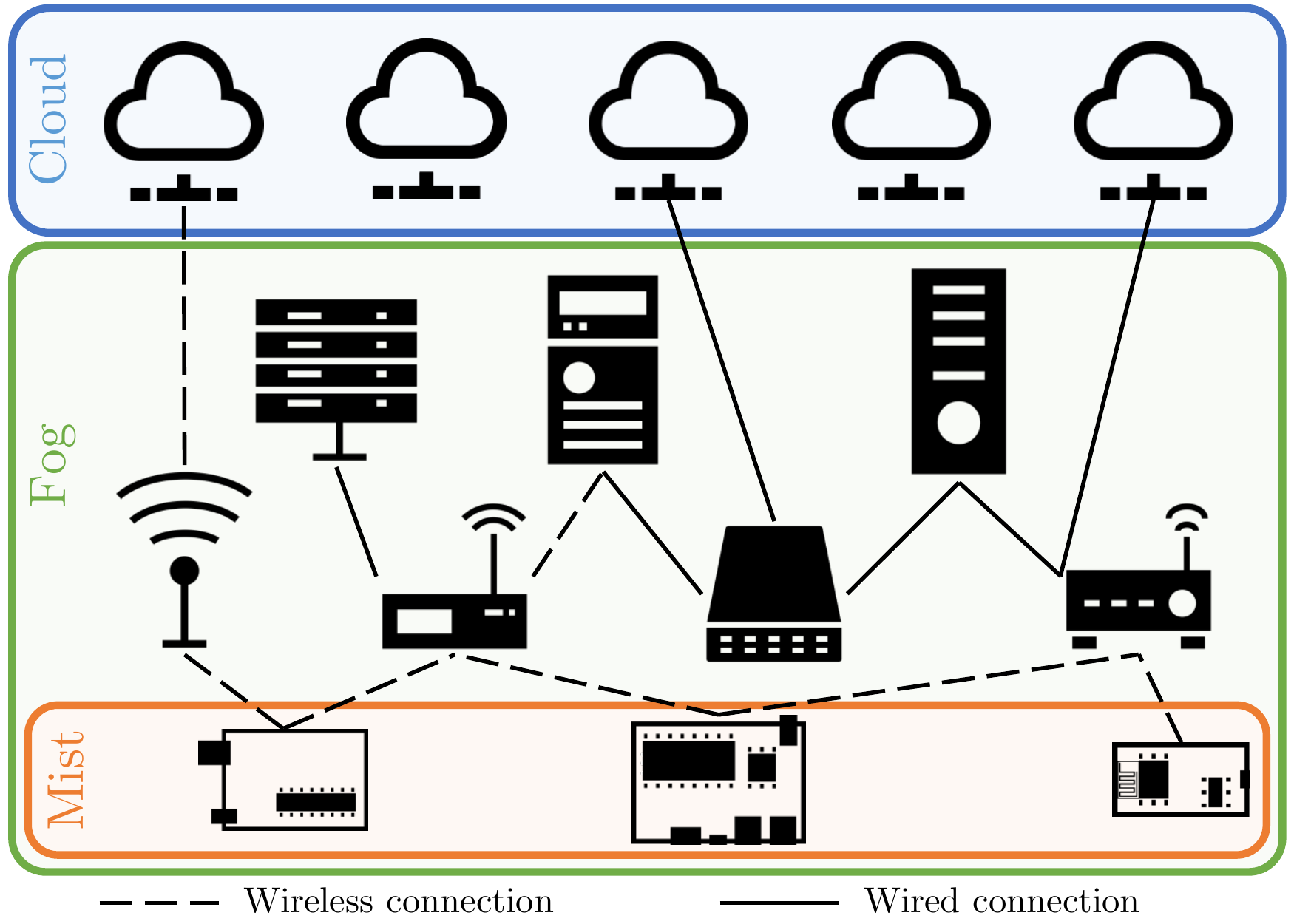}
	\caption{Fog computing hierarchy.}
	\label{fig:fog_hierarchy}
\end{figure*}

In addition to defining fog computing, the NIST Conceptual Model report~\cite{NIST2018-FogComputingConceptualModel} also introduces the concept of \textit{mist computing}\index{mist computing}.
According to them, mist computing can be seen as a ``lightweight and rudimentary form of fog computing that resides at the edge of the network fabric.'' 
They also point out that it ``uses microcomputers and microcontrollers to feed into fog computing nodes and potentially onward towards the centralized (cloud) computing services.''
Figure~\ref{fig:fog_hierarchy} shows an example of the fog computing hierarchy, including how it relates to the cloud and the mist.

The report also states that mist computing nodes are ``more specialized, dedicated nodes that exhibit low computational resources'' which are ``placed even closer to the peripheral devices than the more powerful fog nodes with which they collaborate, often sharing the same locality with the smart end-devices they service.''
However, we have a different proposal for the type of characteristics that should be present in mist nodes, as we consider that resource-constrained IoT devices that are part of the fog hierarchy can be seen as part of the mist.
In this case, the characterization would differ in two main aspects: purpose and placement.

While they expect the purpose of mist nodes to be more specialized and dedicated, we see a valuable opportunity in leveraging resource-constrained devices for custom user code execution, as their proximity to the source of the data will allow them to perform lightweight operations on sensor streams generated by IoT devices, therefore reducing network traffic and latency.
In addition, they characterize the placement of the data source and the mist nodes as separated devices that might share the same locality, while we see that IoT devices usually have both sensors and resources that can be used to service user requests.
This way, some IoT devices would be able to receive and execute the code instead of sending the data to more powerful fog nodes.

With this alternative view on mist computing, we get closer to a setup that is more in line with the advances expected for the IoT, which we consider an important step toward reaching the full potential of this emerging technology.

Given that the mist is part of the fog, when we refer to fog computing in this chapter, we are considering both fog and mist nodes.

\section{Applications}
With the persistent growth in the number of connected IoT devices, they are expected to become continuously smaller, lighter, cheaper, and more energy efficient as technologies progress.
Therefore, in this scenario, understanding constrained devices and being able to work with them will be of paramount importance.

Considering this, we must note that, due to restrictions on resources such as memory and processing power, employing constrained devices may not be appropriate in all situations.
In this section, we explore the characteristics of a few applications and categorize them according to the number of resources they require with the intent of pointing out the cases where constrained devices may present a good alternative.

To this end, we introduce three classes of applications: \textit{Class A0} represents applications well-suited for constrained devices, that is, the ones where only a simple program needs to be executed; \textit{Class A1} has applications that demand a lot of resources, but can make several constrained devices cooperate with each other to process the application in a distributed way; and applications in \textit{Class A9} require considerably more resources and have demands that lead them to be incompatible with constrained devices.

Considering that each particular scenario has its own requirements, there is no one-to-one relationship between all classes of applications presented here and all classes of constrained devices presented in Section~\ref{subsec:constrained_devices}.
Still, we expect that, by providing both software and hardware classifications, we can help developers to identify the type of problem they have at hand, as this can be a good way to start working toward a solution.

Table~\ref{tab:apps} lists some applications in each of these categories, and the following subsections detail one application for each class to illustrate their characteristics.

\begin{table}[htpb]
	\caption{Examples of applications for each class defined in this subsection.}
	\begin{center}
		\begin{tabular}{L{3.5cm}L{3.5cm}L{3.5cm}}
			\toprule
			Class A0					& Class A1					  & Class A9 \\
			\toprule
			Filtering					& Deep learning				  & High-resolution video streaming processing \\
			\hline
			Collect environment data	& Anomaly detection in videos & Self-driving cars \\
			\hline
			Monitor product conditions	& Voice recognition			  & Medical diagnostics \\
			\hline
			Lifelogging		    		& Smart farming	    		  & Scientific applications \\
			\hline
			Smart home		    		& Traffic management		  & Computational Fluid Dynamics  \\
			\bottomrule
		\end{tabular}
	\end{center}
	\label{tab:apps}
\end{table}

\subsection{Class A0}
The main characteristic of Class A0 applications is their simplicity.
They are usually short programs that execute a small number of operations to obtain a specific result, therefore needing fewer resources.
The use of complex libraries is usually taken for granted in many scenarios, but it may present a challenge if an application is to be classified as A0.
However, this does not necessarily mean that Class A0 programs have to rely on specialized or dedicated hardware, as the NIST's conceptual model suggests is the case for mist computing~\cite{NIST2018-FogComputingConceptualModel}.
One way to allow developers to implement a program once and deploy it to many different hardware platforms is the use of Virtual Machines (VMs), as many lean VMs are compatible with constrained devices and thus enable them to execute custom applications.
We refer the reader to Auler et al.~\cite{Auler2016-COISA} for a review on this topic.

\begin{figure}[htpb]
	\centering
	\includegraphics[width=0.9\columnwidth]{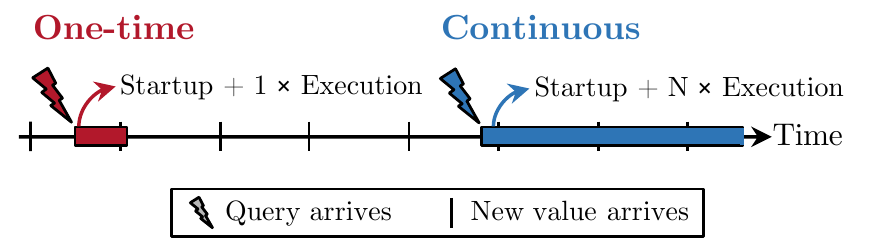}
	\caption{Example of the behavior of one-time and continuous queries.}
	\label{fig:onetime_vs_continuous}
\end{figure}

We can consider a use case where there is an IoT device that is equipped with a capture instrument, like a microphone, or a sensor that periodically measures an environmental variable such as temperature, humidity, or luminosity.
A user (e.g., a data scientist) could be interested in querying this device looking for meaningful information.
To reach this goal, the user would express their query as a program, send it to be executed on the device, and then receive its result.
These requests would be of one of two types: one-time queries, which consider the data of a snapshot taken from the stream at a certain moment in time, are evaluated only once, and return the answer to the user; or continuous queries, which are evaluated continuously as new values arrive and produce answers over time, always reflecting the stream data seen so far~\cite{Babcock2002-ModelsIssuesDataStreamSystems}.
Figure~\ref{fig:onetime_vs_continuous} shows an overall scheme of how an implementation of these types of queries might operate.

Relevant types of queries may be ones that are able to standardize or clean the data (parsing), discard unwanted readings (filtering), or generate knowledge (predicting).
As an example of Class A0 application, we highlight filtering procedures, as these can be simple enough to run on constrained devices and have the potential to drastically decrease the amount of data sent to the cloud, a situation that is poised to become a problem with the large-scale adoption of technologies such as the IoT.

\begin{figure}[htpb]
	\centering
	\includegraphics[width=\textwidth]{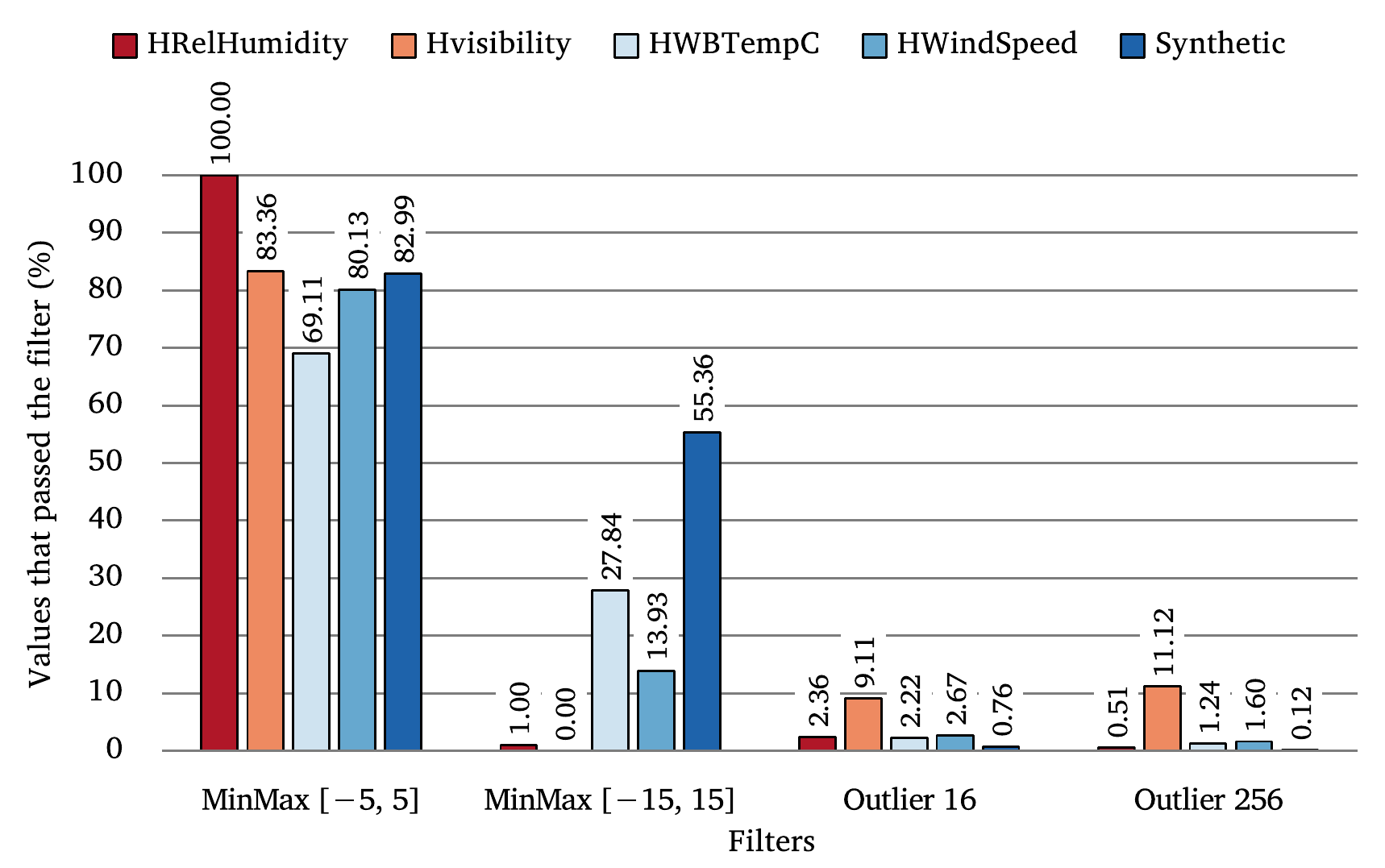}
	\caption{Percentage of values that passed each filter of each dataset.}
	\label{fig:input_output}
\end{figure}

Figure~\ref{fig:input_output} shows an example of the impact that simple filters can have on the number of values that need to be transmitted\index{Internet of Things!climatological data}.
The graph shows the data from five different datasets: four real-world datasets downloaded from the United States National Oceanic and Atmospheric Administration's National Centers for Environmental Information website~\cite{NOAA-NCDC} and an artificial dataset, which represents a stream of sensor readings~\cite{Pisani2017-BeyondFog}.
The four real datasets are each a subset of size 65,536 of the hourly local climatological data collected at Chicago O'Hare International Airport between September 2008 and August 2018.
These datasets are here named HRelHumidity (the relative humidity given to the nearest whole percentage, with values between 16 and 100), HVisibility (the horizontal distance an object can be seen and identified given in whole miles, with values between 0 and 10), HWBTempC (the wet-bulb temperature given in tenths of a degree Celsius, with values between 27.3 and 27.4), and HWindSpeed (the speed of the wind at the time of observation given in miles per hour, with values between 0 and 55).
We call the artificial dataset Synthetic, and it consists of 65,536 values that are random floating-point numbers between $-$84.05 and 85.07.
These values were generated by using the sine function combined with Gaussian noise.

The bars in Figure~\ref{fig:input_output} represent the number of values that are the result of two different continuous filter queries: MinMax, which is a simple filter that only allows numbers outside of a certain range to pass, and Outlier, which is a procedure that executes Tukey's test~\cite{Hoaglin1986-OutlierLabeling} to detect outliers in a window of values.
We chose four instances of these queries: MinMax [$-15$, $15$], which filters out numbers in the [$-15$, $15$] range; MinMax [$-5$, $5$], the same program, but using the [$-5$, $5$] range instead; Outlier 16, which finds outliers in a window of 16 values; and Outlier 256, the same program, but with a window of 256 values instead.
These parameters were selected due to previous work that showed that these four instances filter out very different percentages of stream values~\cite{Pisani2017-BeyondFog}.

The graph shows us that knowing the characteristics of the data with which we are working can make a huge difference in the output of the filter.
For instance, we can look at the HRelHumidity and HVisibility datasets. While filtering out numbers between -5 and 5 on HRelHumidity allows all values to go through due to their range being between 16 and 100, filtering out numbers from -15 to 15 on HVisibility leads to no values passing, given that they are all between 0 and 10.

Still, the most poignant difference of the results is the one between the MinMax and Outlier filters.
While MinMax only looks at absolute values, Outlier identifies abnormal readings, consequently only letting less than 3\% of the values pass through the filter in most cases, even though the value ranges of the datasets differ substantially.

Therefore, we see that executing filters on the constrained IoT devices themselves can significantly decrease the amount of data that would otherwise be sent to the cloud for processing, lowering the network traffic and helping avoid bandwidth congestion.
Furthermore, this type of solution can also present better performance or be more energy-efficient from the point of view of the device than sending all data to the cloud then processing it~\cite{Pisani2019-FogVsCloud,Curado2019}.
It is worth reiterating, however, that these conclusions are only valid for simple programs, and more sophisticated methods would probably not belong to Class A0.

\subsection{Class A1}
Another class is that of applications that demand many resources, but only have access to constrained devices, where it is possible to make the devices cooperate with each other to process the application in a distributed way.
We call this Class A1.

In the scenario of many IoT devices and large amounts of generated data depicted in this chapter so far, we can have two types of working devices: the ones that spend most of their time idle or executing little work~\cite{IoT.Survey.Industries.2014} and the ones that need to process applications that demand a lot of resources.
These devices are likely heterogeneous due to the nature of the application they execute and, consequently, some of them may be constrained devices.
This unveils the need to employ them intelligently in order to make the most of the resources we have at hand.

For example, deep learning techniques have been successfully applied to the interpretation of the kind of data generated by the IoT~\cite{Alex.2012, DeepX.2016, Pinkesh.2017}.
However, methods such as Convolutional Neural Networks (CNNs)\index{Internet of Things!Convolutional Neural Network} require a lot of computational power, a demand that is aggravated in a system with resource-limited devices.
The majority of previous works focus on the optimization of the training phase of neural networks~\cite{performance.modeling.2015, SINGA.platform.2015, Ako.2016}, as it presents a higher computational cost, but in the case of constrained devices, even the inference phase may be too computationally intensive.
As the inference of neural networks is a suitable application to process the data generated by the IoT, we may want to execute it on constrained devices.
Therefore, we need to optimize this process in order to comply with the real-time response or inference-rate requirements that the application may have.
Besides that, current deep neural networks may be too large for the limited amount of memory available on these devices, so it would not be possible to deploy them on a single constrained IoT device.

\begin{figure}[htpb]
	\centering
	\includegraphics[width=\textwidth]{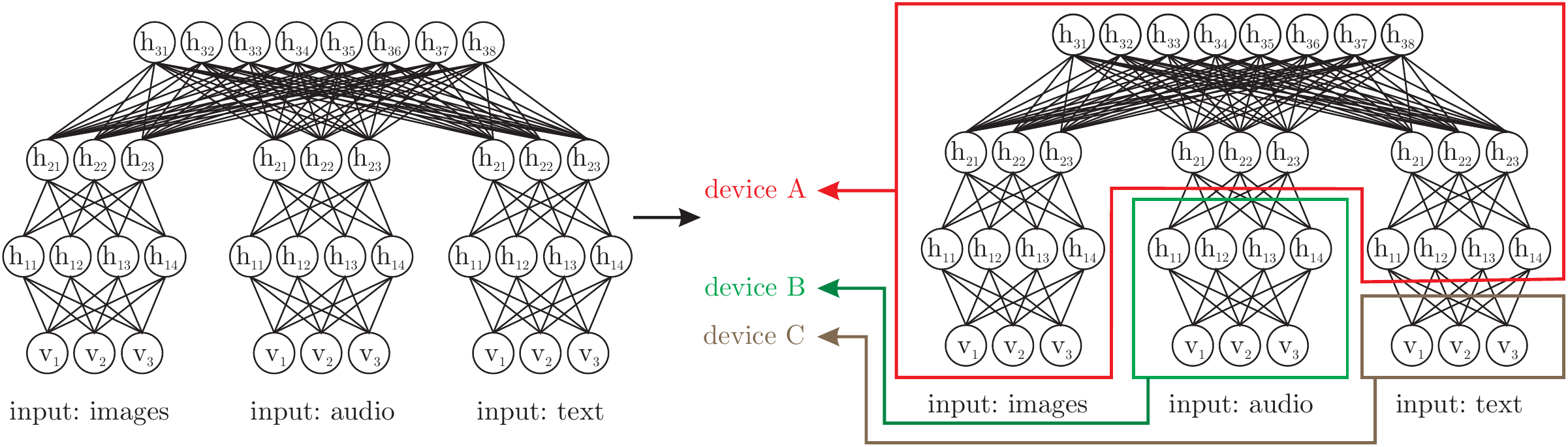}
	\caption{How a neural network may be partitioned to be executed distributedly on three constrained devices.}
	\label{fig:distributed}
\end{figure}

In the Class A1 of our application classification, we want to execute an application that demands a great amount of resources but only has access to constrained devices, so, in order to enable the execution, we need to partition the neural network model to distribute it on multiple devices.
Figure~\ref{fig:distributed} shows an example of how a neural network may be partitioned to be executed on three different devices in a distributed way, considering that each device provides different amounts of memory, computational power, and energy life.
There are already frameworks that can execute the computation of neural networks in a distributed manner such as TensorFlow~\cite{Tensorflow2.2016}, DIANNE~\cite{DIANNE.2015}, and DeepX~\cite{DeepX.2016}.
DIANNE and DeepX are specific for the IoT and focus on the inference phase, but only offer the option to partition the neural network into its layers.
TensorFlow only presents this partitioning restriction if the user wants to take advantage of its implemented functions.

\subsubsection*{\textbf{Machine learning frameworks and partitioning algorithms}}

Although the current frameworks mainly offer only one simple option to partition neural networks for distributed execution, different partitioning approaches are possible; nonetheless, some of them provide suboptimal performances according to the application requirements~\cite{sbac-pad.2018}.
In addition to the per-layer partitioning offered by these tools, we can manually partition the neural network model, an approach that may be time-consuming, error-prone, suboptimal, and whose performance depends on the expertise of the user, in terms of both machine learning and the partitioning itself.
A third option is to use a general-purpose partitioning algorithm to try to achieve at least a near-optimal result.
As the last possibility, we can look at the details of the neural network to develop a specific partitioning tool, an idea that we are going to explore in Section~\ref{sec:challenges_opportunities}.
In this subsubsection, we discuss machine learning tools and general-purpose partitioning tools.

TensorFlow~\cite{Tensorflow2.2016} is a machine learning framework created by Google that is able to distribute both the training and inference of neural networks on heterogeneous systems, from mobile devices to large clusters.
To use it, the developer needs to partition the neural network manually, which includes the assignment to GPUs.
They are also allowed to partition each neural network operation, but this approach prevents them from taking advantage of TensorFlow's implemented functions.
To use these functions, the partitioning must be made into layers, as is required by other machine learning tools.
Furthermore, we point out that TensorFlow was designed to speed up the training of neural networks, thus it does not consider important challenges for constrained IoT devices such as energy, memory, computation, and communication constraints.

De Coninck et al.~\cite{DIANNE.2015} implemented a framework called DIANNE to distribute neural networks on the IoT, which can model, train, and evaluate neural networks over multiple devices.
This tool optimizes the inference of each sample at a time for streaming on IoT devices, requiring the user to manually partition the model in a per-layer approach.

DeepX~\cite{DeepX.2016} was presented by Lane et al. and employs three approaches to enable the execution of neural networks in the IoT.
First, it compresses the neural network layers of the model so that the memory and computation requirements decrease.
After that, it may offload some layers of the neural network to the cloud for execution.
Finally, at each inference request, DeepX is able to decide between its CPU or GPU and the cloud, performing a simple dynamic scheduling between its only node with multiple processors and the cloud.
The possible drawbacks in this strategy are the fact that their tool may be too computationally intensive to be executed on devices that are more resource-constrained than smartphones and that the model can only be partitioned into layers.
Moreover, DeepX cannot make use of resources that are available in the same environment, as it only looks to one node and the cloud.

SCOTCH~\cite{SCOTCH.2009} is a software package designed for general-purpose graph partitioning and static mapping.
It partitions a graph attempting to balance the computational load and reduce communication costs.
Given that this tool is not specific for constrained devices, it does not consider memory constraints.
However, it accounts for redundant edges, that is, multiple edges that come from one vertex and represent the transfer of the same data to the same partition, a common situation in machine learning.

Kernighan and Lin proposed a graph partitioning algorithm~\cite{Kernighan.1970} that was widely applied to distributed systems~\cite{KLapp1.2012, KLapp2.2012, KLapp3.2014}.
It is a heuristic where a graph representing the computation is first divided into random partitions.
Then, the algorithm calculates the communication cost for this initial partitioning and analyzes the gain or loss in the communication cost that would result from swapping a pair of vertices where each belongs to a different partition.
After analyzing every possible pair, it chooses the one that provides that largest gain, even if it maintains or increases the current communication cost.
Once it selects a pair for swapping, the vertices are locked and cannot be swapped again until every pair in the partitioning is chosen.
When this happens, one epoch of the algorithm passed and the whole process may be repeated while it is still obtaining improvements.
With this heuristic, the authors believe that it is possible to reach a near-optimal partitioning.

METIS~\cite{metis.1998} is an open-source software package for partitioning large graphs and meshes and computing orderings of sparse matrices developed at the University of Minnesota.
For the partitioning, it employs the multilevel graph partitioning paradigm, which consists in successively grouping the graph vertices based on their adjacency until the graph is composed of only hundreds of vertices.
It then applies some partitioning algorithm such as Kernighan and Lin to this small graph and gradually returns to the original graph, performing refinement steps using the vertices at the edge of the partitions during this process.

All the tools presented here can be useful for the distributed execution of neural networks, although the machine learning frameworks lack an automatic, flexible partitioning while the general-purpose partitioning tools lack an appropriate treatment of memory restrictions, redundant edges, and other specificities that are relevant for machine learning applications executed on constrained IoT devices.
These gaps raise some challenges and opportunities that we detail in Section~\ref{sec:challenges_opportunities}.

\subsection{Class A9}
Applications in \textit{Class A9} require considerably more resources, thus needing devices that are more powerful to support them and fulfill their requirements.
We illustrate this class with high-resolution video processing and transmission~\cite{Shielding2016,TowardsH2652016}.

The availability of low-cost, low-power heterogeneous multimedia devices has gained traction with the advent of constrained-device networks for smart city environments.
These multimedia devices are envisioned as small-sized objects equipped with limited power, which they have to utilize efficiently to increase their network lifetime.
However, the delivery of multimedia data should still meet the communication constraints (i.e., delay and jitter), which require higher bandwidth and efficient communication mechanisms. 

In order to process the video stream\index{Internet of Things!video stream}, several pre-transmission procedures are executed on the multimedia device that is capturing it, for example, media encoding, decoding, storage, estimation, entropy coding, etc.
These processes are computationally complex and consume a significant amount of resources, thus fitting in the Class A9 of our classification.

However, it is possible to use various promising solutions proposed for efficient multimedia communication to facilitate part of the multimedia acquisition and processing on the multimedia device.
For instance, we can use distributed video coding to partition the complexity between the encoder and decoder, turning this part of the solution into a Class A1 application.
In particular, most of the literature allows encoding with a very low complexity when compared to that of decoding.
Applications such as wireless video surveillance use mobile video cameras with limited computing resources and may benefit from this lightweight encoding with high compression efficiency.

Moreover, there have been studies about compressive sensing with the intent of reducing the complexity of the encoding process.
Considering the energy limitations of the devices, compressive sensing-based techniques can improve their life cycle.
We can use such methods in constrained devices to enable video streaming operations to be performed even in cases where there are energy limitations.
It is possible to notice that there is a trade-off between the communication and the processing costs at each level of the user experience, and the developer of Class A9 applications can analyze the characteristics of the scenario with which they are working to try to find any compromises that may allow them to use constrained devices.

Low-power IEEE 802.11 devices supporting much higher data rates are being designed and are expected to be capable of operating services for video streaming provisioning.
This could lead to this type of application becoming compatible with constrained devices in the future.
The routing protocol for low power and lossy networks (RPL) in the current IoT communication stack is flexible and can be adapted to operate in an energy-efficient way considering different application requirements.
Still, the current MAC and PHY layers proposed for the IoT (e.g., ZigBee) only support a theoretical data rate of \SI{250}{kbps}, which is far less than a typical multimedia-based application requires.
Nevertheless, these protocols are still adopted for the IoT communication stack due to their energy-efficient operation.

With the ongoing advances in technology, we anticipate that video streaming devices, such as cameras or microphones, will become globally accessible by unique IP addresses in a similar way to computers and other networking devices currently connected to the Internet.
Furthermore, some computing capability may be embedded in these multimedia devices to make them smart enough to perceive system and service requirements and to trigger actions on their own.
Therefore, we could use heterogeneous multimedia devices to acquire and process multimedia content from the physical environment (e.g., audio, video, and images), thus being able to execute this type of Class A9 application.

\section{Future Scenarios for Fog Computing and the IoT}
Having in mind the concepts of fog computing and IoT and the types of applications that can be executed on constrained devices, we can now think about what we expect to see in a future where these technologies are combined.
In order to do that, we will start by depicting a few speculative scenarios for future applications.

\begin{figure}[htpb]
	\centering
	\includegraphics[width=0.7\textwidth]{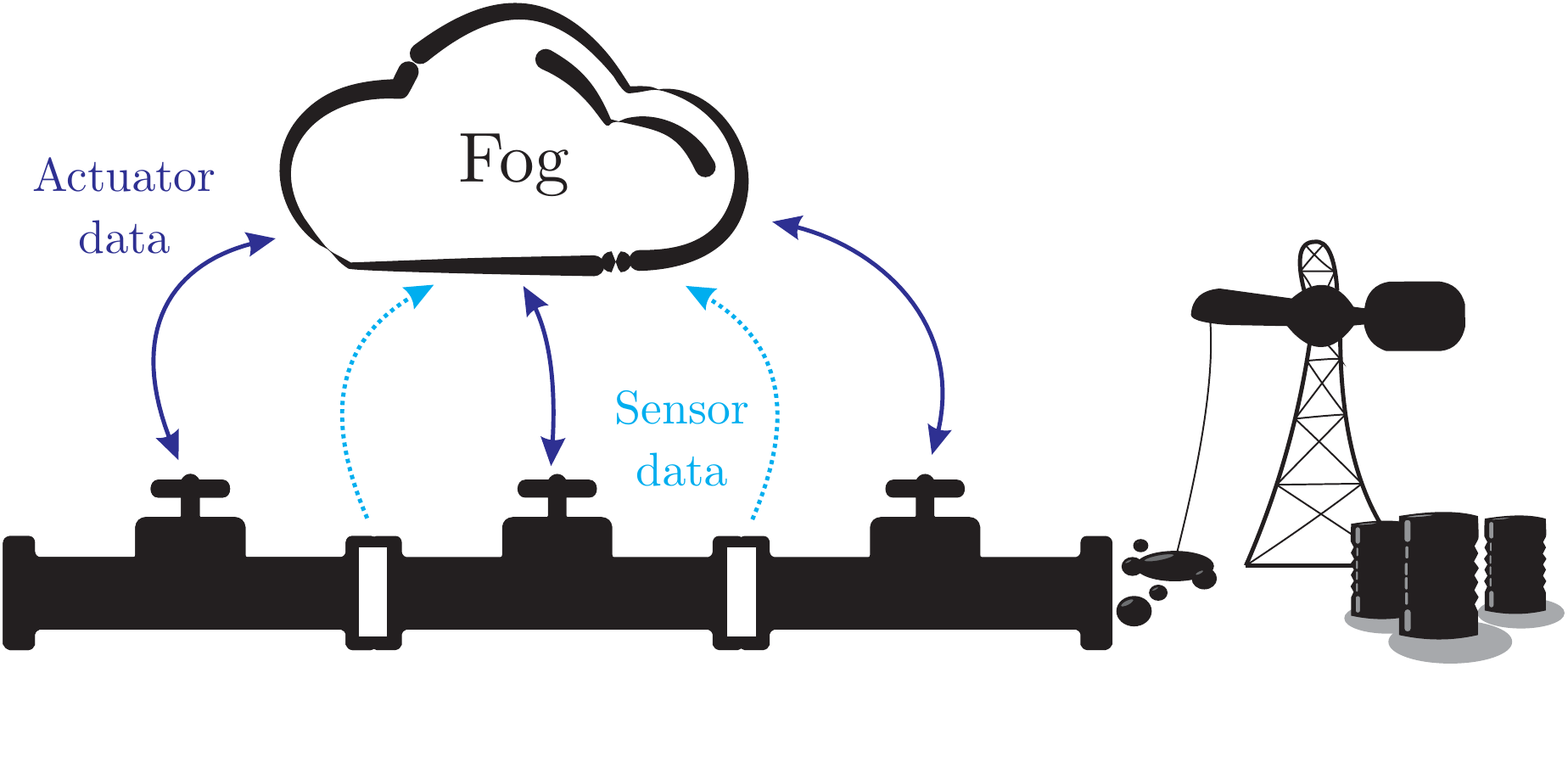}
	\caption{Sensors distributed along an oil pipe supervising its structural health.}
	\label{fig:ex1-present}
\end{figure}

First, there is the case of oil pipes, which can use pressure and flow sensors placed every several meters to detect leaks and control valves~(Figure~\ref{fig:ex1-present}).
While it is useful to be able to generate periodic reports about the collected data, it can be costly to send all sensor readings to the cloud~(using perhaps an expensive satellite link) and the response time of hundreds of milliseconds may not be fast enough to close the valves and prevent massive oil spills in time.

\begin{figure}[htpb]
	\centering
	\includegraphics[width=0.7\textwidth]{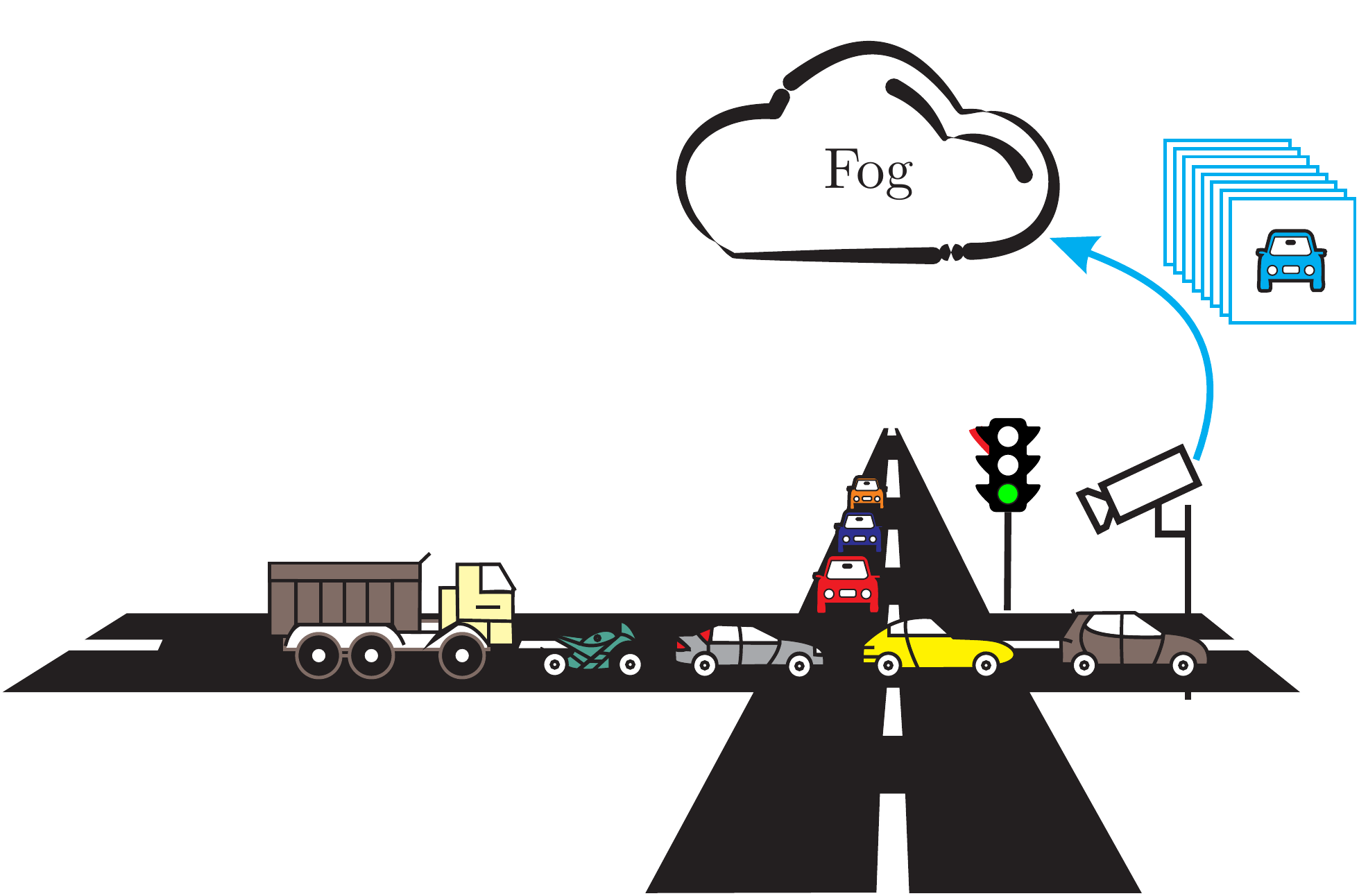}
	\caption{Smart traffic camera detecting road conditions.}
	\label{fig:ex2-present}
\end{figure}

The second instance is the use of smart cameras to automatically monitor traffic and detect incidents, identify construction zones and other road impediments, and find problem areas where there are issues such as frequent slowdowns~(Figure~\ref{fig:ex2-present}).
While these data may be sent to the cloud to be used for analytics and to help inform people about traffic conditions in certain areas, it is important that the system continues working even when the connection to the cloud is down so that first responders may be called as soon as possible in case of emergencies.

\begin{figure}[htpb]
	\centering
	\includegraphics[width=0.7\textwidth]{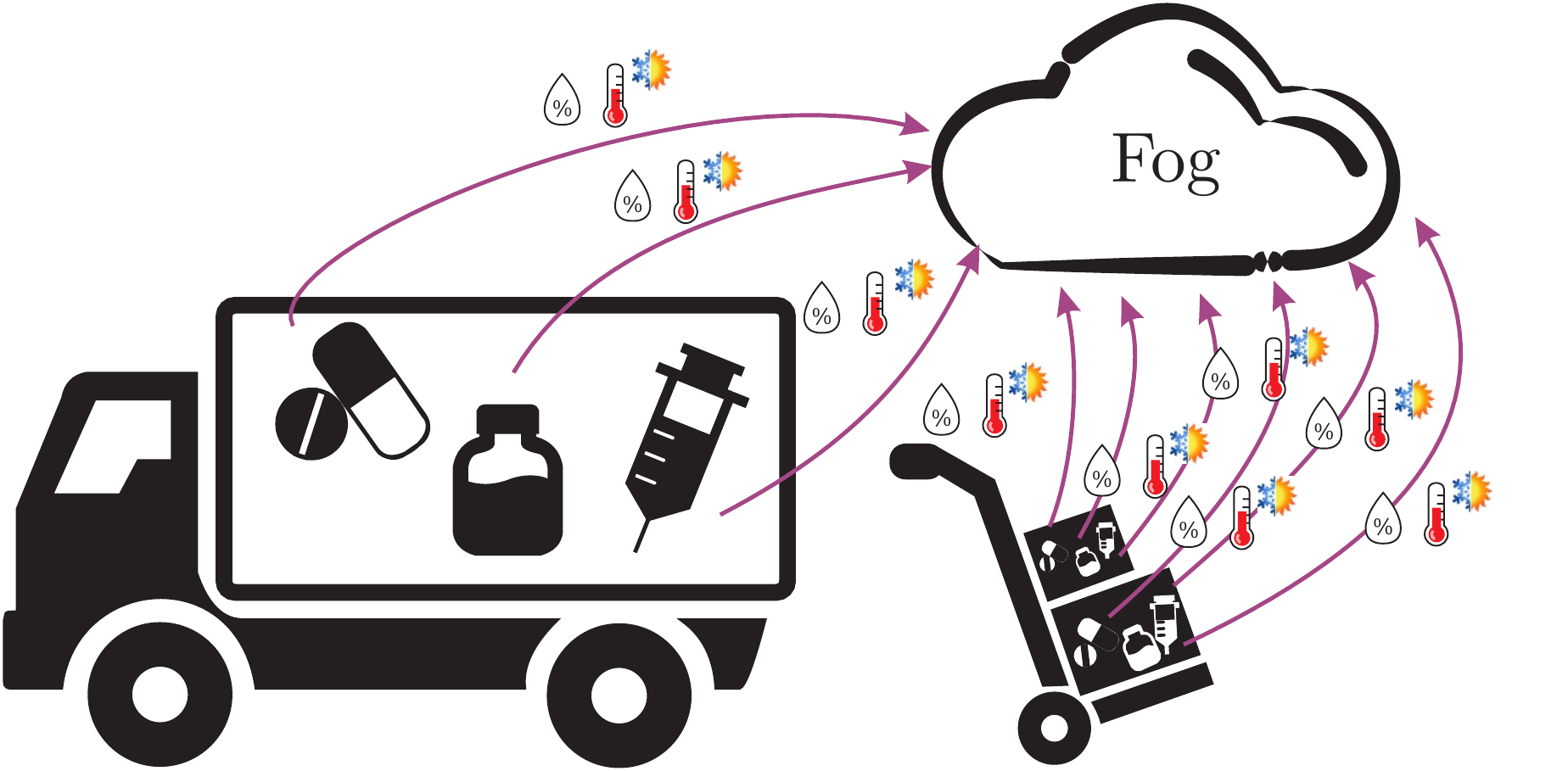}
	\caption{Sensors tracking the environmental status for medicine packages.}
	\label{fig:ex3-present}
\end{figure}

A third example is the transportation of medicine and vaccines, as there is a risk of compromising the efficacy and safety of these products if they are exposed to excessive heat, cold, or humidity (Figure~\ref{fig:ex3-present}).
By placing temperature and humidity sensors in each package, it would be possible to guarantee that they are being stored under proper conditions. However, it may not be interesting to save every single data point collected, given that the focus would be only in the situations where values are outside of the expected boundaries.

When we look at these applications, they seem like a natural fit for the anticipated short-term progression of fog computing, and it is possible to expect this type of cases to become a part of our lives sometime within the next decade.
However, what if we continue this exercise, and turn our attention to a future that is even further ahead? Thirty, maybe fifty, years from now?

\begin{figure}[htpb]
	\centering
	\includegraphics[width=0.7\textwidth]{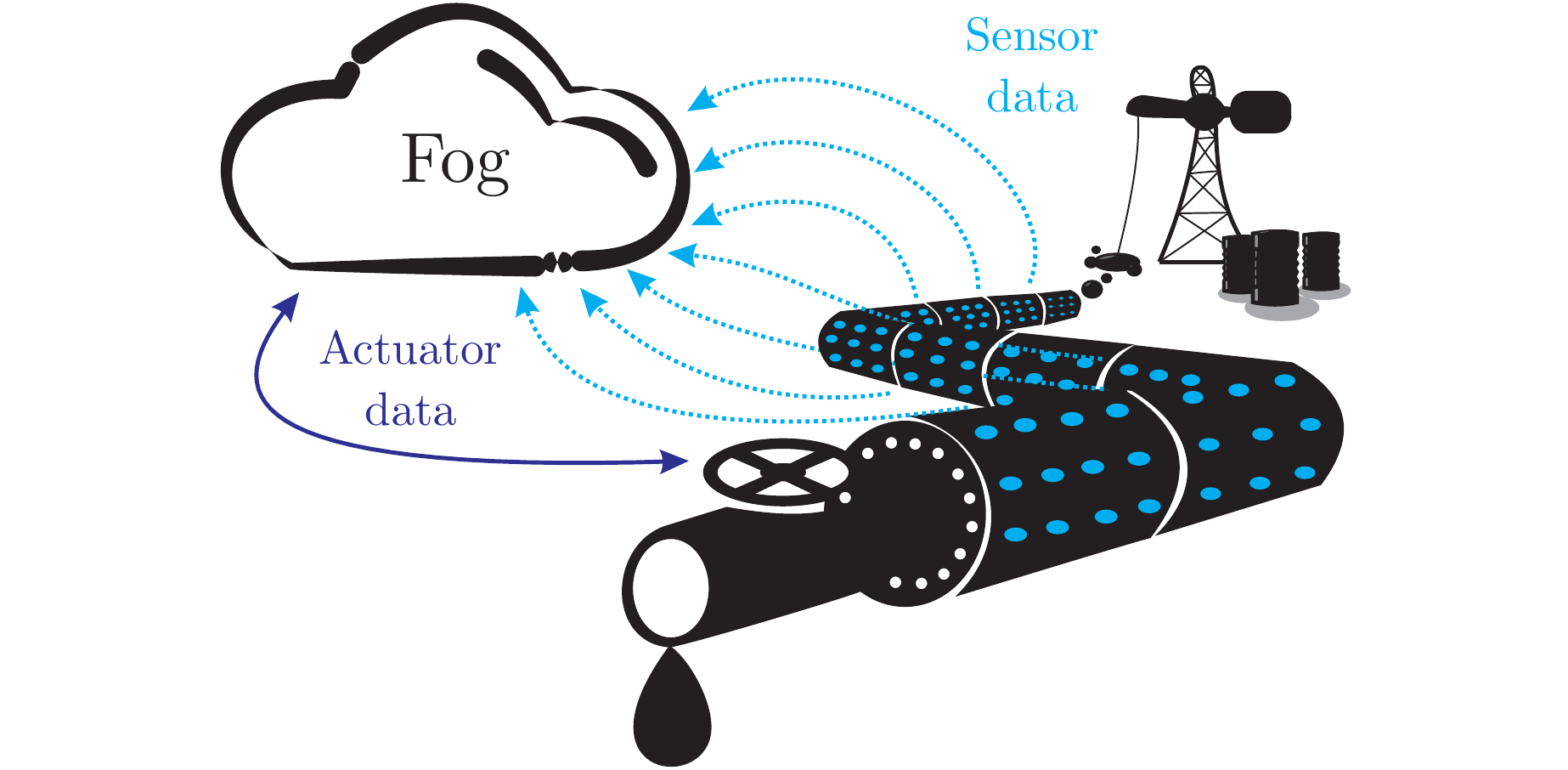}
	\caption{Sensor mesh covering an oil pipe monitoring its integrity.}
	\label{fig:ex1-future}
\end{figure}

By that time, technological advancements and cost reduction should enable the deployment of many more sensors than what we described in our scenarios.
For example, instead of placing sensors on every several meters of an oil pipe, it may be possible to deploy a sensor mesh that covers the whole structure~(Figure~\ref{fig:ex1-future}), allowing us to not only find leaks but also to monitor the pipe's structural integrity and predict failures before they even start interfering with the system.

\begin{figure}[htpb]
	\centering
	\includegraphics[width=0.7\textwidth]{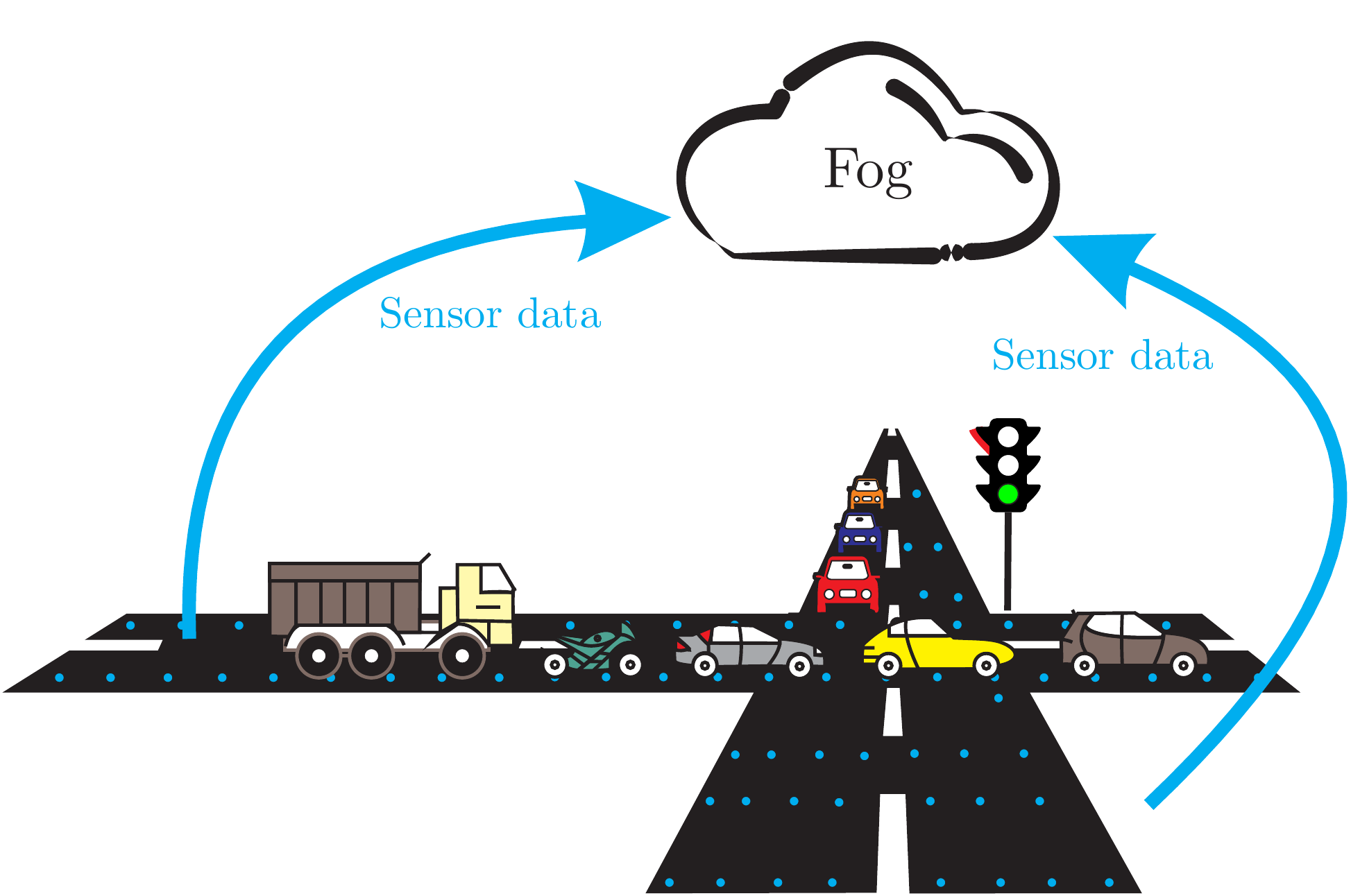}
	\caption{Sensors embedded in the asphalt checking for road impediments.}
	\label{fig:ex2-future}
\end{figure}

Rather than using expensive smart cameras to detect road conditions, cheap sensors could be mixed with the asphalt~(Figure~\ref{fig:ex2-future}) and closely detect the presence of road impediments and traffic conditions.

\begin{figure}[htpb]
	\centering
	\includegraphics[width=0.7\textwidth]{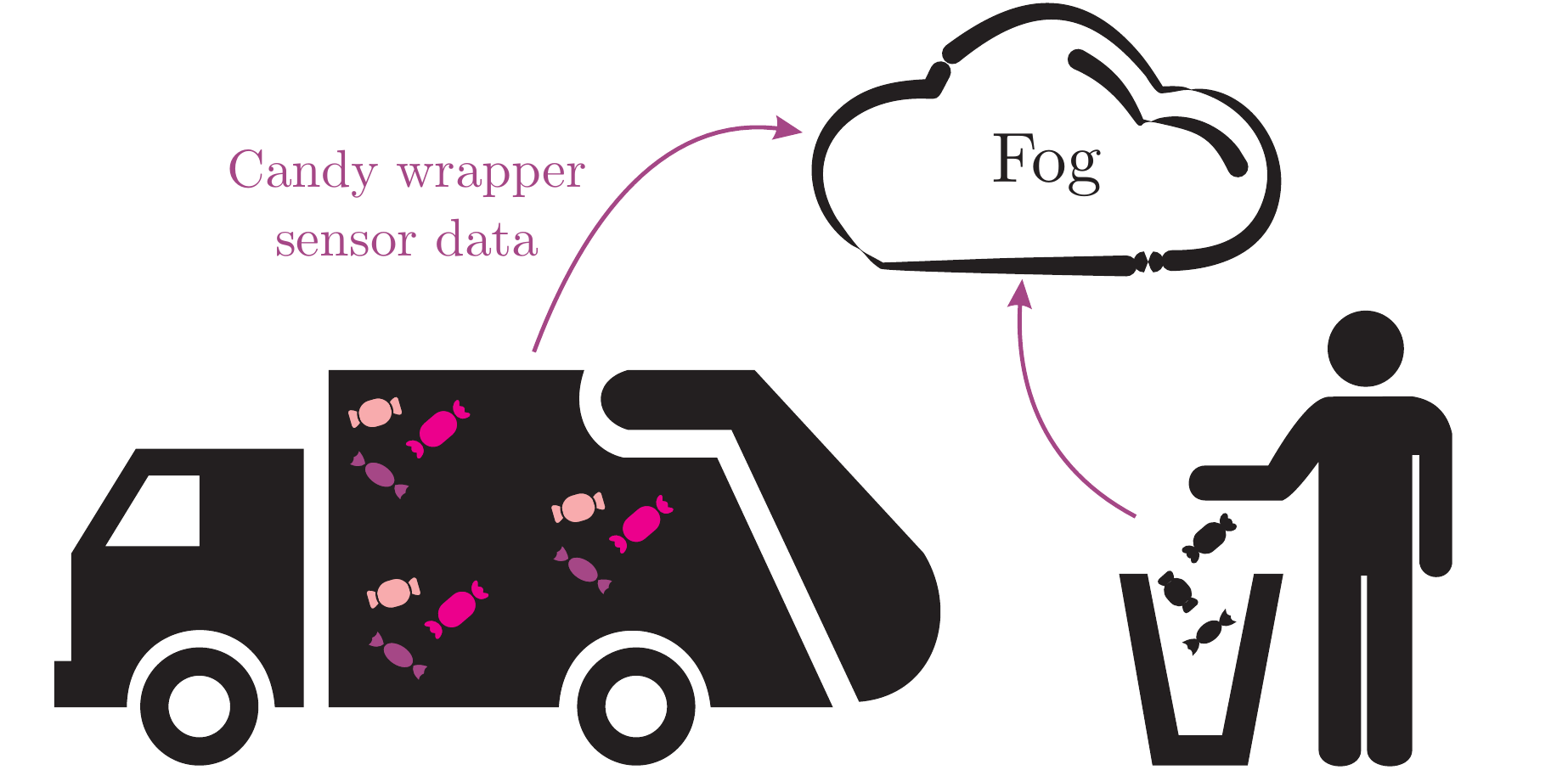}
	\caption{Sensors informing the location of candy wrappers.}
	\label{fig:ex3-future}
\end{figure}

It may be even possible to go beyond placing sensors only on important containers such as medicine packages and have them also installed on simple candy wrappers~(Figure~\ref{fig:ex3-future}) to help with transportation logistics, inventory, and littering detection.

Some of these new scenarios may sound a little excessive or maybe outright infeasible.
However, the same would be true about some of the technologies currently available if they had been described several years in the past.
Consider the printing press, for instance.
A few centuries ago, nobody would have expected that one day it would be used to write nutritional information on the paper used to wrap candy, but now it is an industry standard and very commonplace.
Moreover, with the accelerated rhythm in which technology is currently evolving, we will not have to wait hundreds of years to see technological breakthroughs become a part of our daily routine; this future is only decades away.

Looking at all of these possibilities, one question that arises is what is possible to do now to progress toward the future we described.
Although we can see that both academia and industry have already taken the first steps to develop an infrastructure that will enable the IoT, we are still far from the scenario where this technology is truly pervasive.
Considering that in this future there will be an immense number of cheap, small, low-energy, and lightweight devices, with challenging limitations such as small memory and low processing capacity, efforts to integrate them as nodes into the fog hierarchy such as what is proposed by mist computing sound like a promising direction.
We will discuss other challenges and opportunities related to this question in Section~\ref{sec:challenges_opportunities}.

\section{Challenges and Opportunities}
\label{sec:challenges_opportunities}
\begin{figure*}[htpb]
	\centering
	\includegraphics[width=\textwidth]{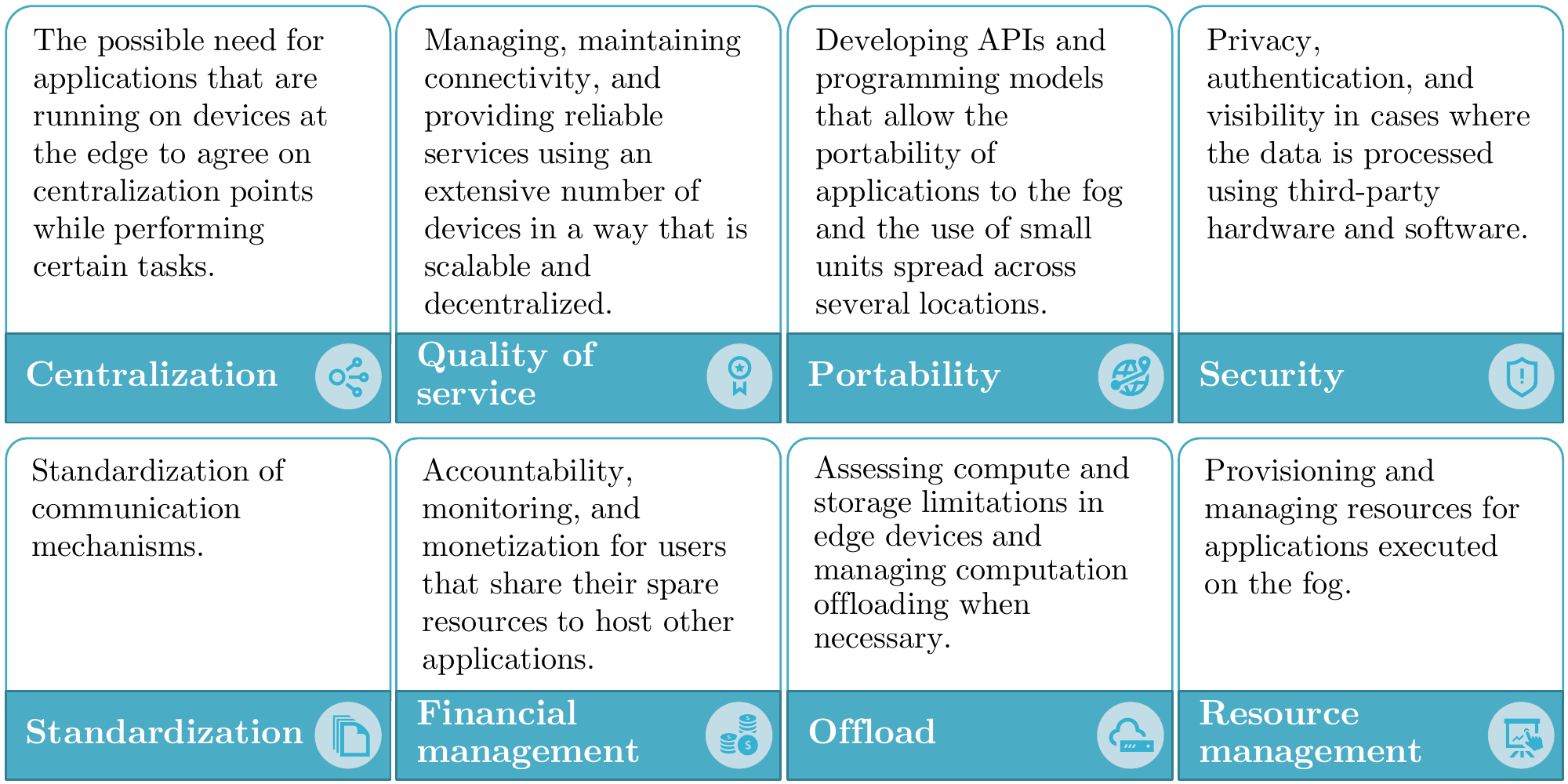}
	\caption{Challenges for fog computing.}
	\label{fig:fog_challenges}
\end{figure*}

Along with all the possibilities brought by fog computing and the IoT comes a series of challenges that stand in the way of their full realization (featured in Figure~\ref{fig:fog_challenges})~\cite{Vaquero2014-FindingYourWayInTheFog,Yi2015-SurveyFog}.
In this section, we further discuss these challenges and opportunities while considering the application classes presented in Section~\ref{subsec:constrained_devices}.

\begin{itemize}
    \setlength\itemsep{0.5em}
	\item \textit{Centralization:} this challenge is valid for applications that are running on devices at the edge and have to agree on centralization points to perform certain tasks.
	Some operations can have data-intensive and compute-intensive parts, and defining these two phases well helps to determine whether the direct interaction with the fog/cloud will be possible or cost-prohibitive.

 	\item \textit{Quality of Service (QoS):} with the increase of IoT devices being covered by fog nodes, supporting surging data traffic with adequate QoS levels is an important objective in the design of these applications.
 	This way, managing, maintaining connectivity, and providing reliable services using an extensive number of devices in a way that is scalable and decentralized is challenging.

	\item \textit{Portability:} constrained devices must also allow for code portability, with developers implementing a program only once and deploying it to many different hardware platforms.
	The development of APIs and programming models that allow the portability of IoT-to-Fog applications remains an open challenge.

    \item \textit{Security:} the system development without due care to data protection results in environments that can be extremely vulnerable.
    A secure IoT-to-Fog system involves the security of data in transit in the network as well as within the devices themselves.
    Three elements are important when the data processing uses third-party hardware and software: privacy, authentication, and visibility.

	\item \textit{Standardization:} standardization of communication mechanisms would allow better configuration and communication among devices and management entities across all levels of the IoT-Fog-Cloud hierarchy.
	Standardization efforts comprising industry and academia interactions, as exemplified by the Industrial Internet Consortium, are desired to establish standards in this scenario.

	\item \textit{Financial management:} in common fog systems, there are three incentive schemes based on monetary, reciprocity, and reputation rewards for users to share their spare features to host other applications.
	Multi-party sharing of devices requires mechanisms that are transparent and agreed upon to enable accountability, monitoring, and monetization of shared resources in a way that encourages participation.

	\item \textit{Offload:} one challenge in offloading computation from one device to another is how to handle the different IoT application requirements.
	For example, we must evaluate the resources needed by an application to decide how to partition its tasks in real time among devices with different compute and storage limitations.
	Considering that IoT environments are usually highly heterogeneous, this can lead to complicated offloading decisions.

	\item \textit{Resource management:} provisioning and managing billions of devices that run one or more services is a complex task.
	In order to do that, novel resource management mechanisms need to be developed through machine-to-machine and device-to-device communications, which can improve the QoS without jeopardizing the application services.
\end{itemize}

\subsection{Tackling IoT Challenges with Fog Computing}
\label{subsec:iot_challenges}
\begin{figure}[htpb]
	\centering
	\includegraphics[width=0.6\textwidth]{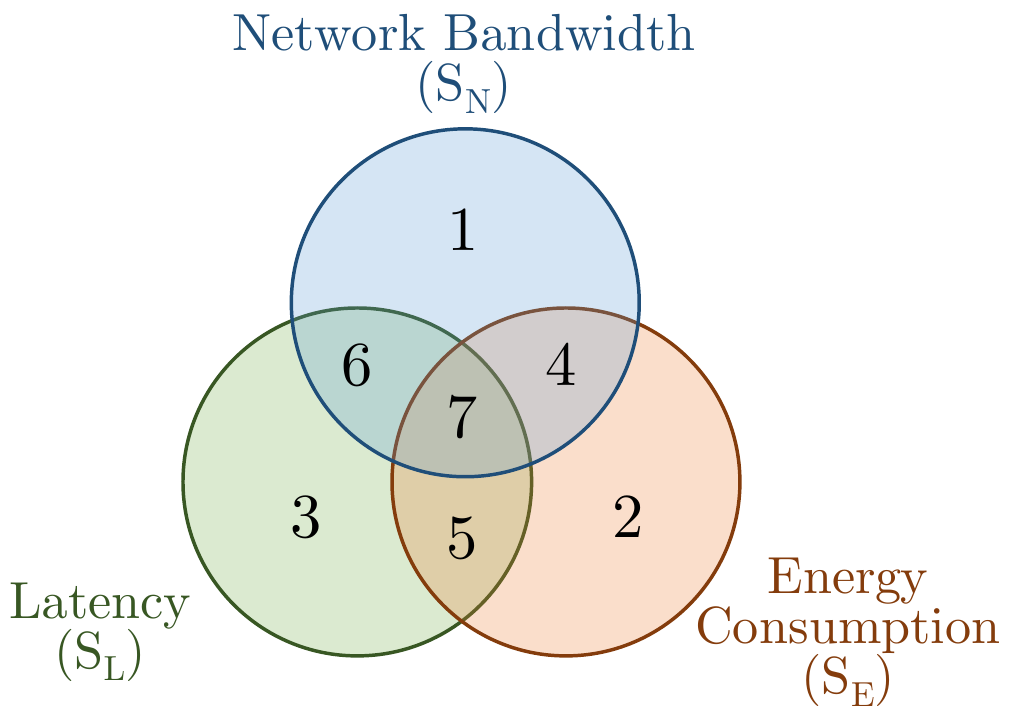}
	\caption{Challenges for the Internet of Things.}
	\label{fig:iot_challenges}
\end{figure}

Like any new technology, the Internet of Things comes with a series of challenges that must be addressed for it to realize its full potential.
In this chapter, we focus on three characteristics that are directly linked with the execution of computation on IoT devices: network bandwidth, energy consumption, and latency.
Figure~\ref{fig:iot_challenges} shows a Venn diagram where each set represents the applications for which a certain characteristic is a challenge. 

The \emph{network bandwidth} set~($S_{N}$) represents applications where the amount of data produced by the devices is so large that sending it through the network becomes too costly or even impossible, which may also be the case of setups with very slow or intermittent network connections.
The \emph{energy consumption} set~($S_{E}$) includes applications where there is a constrained power budget and therefore low energy usage is very important, which is the case of devices that, for example, are battery-operated or depend on limited energy sources such as solar power.
Finally, the \emph{latency} set~($S_{L}$) represents problems where there is a strong time constraint for the result to be produced, sometimes even a (near) real-time application.

The combination of cloud and fog computing will offer different resource allocation possibilities that will enable applications with combined requirements, such as low latency and low energy consumption (region 5 in the diagram), to be handled more efficiently than using only cloud computing or edge computing alone.
Resource management and allocation for each of the seven combinations of application challenges shown in Figure~\ref{fig:iot_challenges} may require different approaches.
The identification of these requirements, as well as the development and evaluation of resource allocation policies for each application, are currently being studied.
However, the question of whether we need a single adaptive mechanism or distinct specific mechanisms for addressing those challenges remains open.

\subsection{Challenges and Opportunities in Class A1 Applications}
When considering the execution of neural networks on constrained IoT devices, the user can distribute the computation on more than one device to enable execution on memory-constrained devices and/or to increase the application performance.
In this situation, memory is a key restriction for any partitioning, as the application cannot be deployed if the partitioning does not respect this constraint.
Therefore, a tool that considers memory as a restriction instead of only attempting to balance it is an opportunity in the deployment of machine learning tools for constrained IoT devices.

Some partitionings are better than others for each objective function that can be employed to comply with the application requirements, so a proper partitioning algorithm should be used to find the most suitable partitionings.
Thus, the development of a tool that offers an automatic and efficient partitioning is both a challenge and an opportunity in this scenario.
Moreover, one of the goals of this partitioning algorithm should be reducing communication, as this reduces the amount of power consumed in radio operations and the amount of interference on the wireless medium.
The simple approaches of current machine learning frameworks such as TensorFlow, DIANNE, and DeepX may increase the required communication between IoT devices~\cite{sbac-pad.2018}, so another opportunity is to integrate the automatic partitioning algorithm in the machine learning framework in order to provide a complete tool.

Looking from the perspective of the IoT, one challenge/opportunity in developing such an automatic partitioning tool is the possibility of eliminating some unnecessary devices from the partitioning in case they harm the objective function.
This necessity comes from the fact that an IoT framework may search for the connected devices in a certain environment, but not all of them may be considered in the partitioning.
Furthermore, as often is the case, the partitioning should be performed into more than two partitions, and some approaches solve this problem by recursively finding two-way partitionings.
However, this strategy may fall into local minima and to improve this result is both a challenge and an opportunity.

A common approach to partitioning the graph is to try to balance vertices in the hope of achieving a good partitioning; however, the current best results focus on maximizing or minimizing the problem objective function instead~\cite{sbac-pad.2018}.
This leads to another challenge and opportunity, as considering several objective functions, one for each application requirement, may enhance the execution of neural networks in the IoT.
These functions can be the maximization of the inference rate or the battery life of the devices, or the minimization of latency, energy consumption, or communication.
Having more than one objective function forming a multi-objective problem as well as deploying each of these objective functions are also challenges and opportunities.
Finally, another challenge and opportunity arise in finding good partitionings for heterogeneous environments such as the IoT considering computational power, energy, memory, and communication.

\subsection{Challenges and Opportunities in Class A9 Applications}
Typical processing of the videos captured by persistent surveillance sensors involves extracting and tracking entities (e.g., vehicles, people, and packages) over time from the raw image data.
Full Motion Video with 30 to 60 frames per second at High Definition resolution and Wide-Area Motion Imagery formats make the data too large to be easily transmitted.
Therefore, this context will be challenging yet necessary, given that the constrained devices have limited computation and Internet access bandwidth.

The distribution of Class A9 application data from the edge to the computing infrastructure hierarchy usually involves a trade-off between increasing network traffic and delay in order to obtain higher computing capacity.
Nevertheless, the higher a workload and its dataset travels in the IoT-Fog-Cloud hierarchy, the more data is transmitted through the links of the network, as more hops are traversed.
Therefore, this application class brings challenges on how to process data throughout the infrastructure layers without hampering application efficiency and without overloading network links in the computing hierarchy.
Approaches that are concerned with the distribution of tasks along with the computing layers, such as the management of Vertical Workflows~\cite{rana2018vertical}, are still under development and present both a challenge and an opportunity.

\subsection{Network Communication Stack Considerations}
While current challenges in IoT network simulations seek to minimize the bandwidth consumption and maximize network throughput, there is a lot less work related to the communication stack using real testbeds.
In fact, the low-power communication protocols (e.g., 6LoWPAN and CoAP for the IP and Application layers, respectively) address only constrained devices in Class A9 whereas there are challenges to be tackled in Classes A0 and A1 as well.

In such networks, the routing protocols have to be highly flexible and adaptive to network changes, so it is both an opportunity and a challenge for routing algorithms to consider energy-efficient routes for data traffic.
Nonetheless, IoT communication poses demanding requirements, which is why some protocols designed for the IoT may not be feasible for constrained devices, as these are expected to be low-cost and have limited resources.

There are other opportunities to explore, such as setting up different configurations in the communication between IoT devices (e.g., data packet size, channel communication, and signal power strength), which can alleviate the burden on memory.
However, changing physical communication protocols according to device technologies may also be a challenge, considering that it can affect interoperability.
For example, Zigbee and Wi-Fi share the same frequency range but do not communicate with each other.

Lastly, the lack of real multi-hop networks is also a challenge/opportunity, given that constrained devices may need to communicate with applications through other nodes in real-world scenarios where the PHY/MAC layers have different behaviors when it comes to multi-hop protocols.

\section{Conclusion}
In this chapter, we gave an overview of the concepts of constrained devices, the Internet of Things, as well as fog and mist computing.
We highlighted the importance of constrained devices in the long-term future of the IoT and discussed the implications of leveraging restricted hardware in this possibility-filled scenario.

We pointed out that, to pave the way for a world where the full potential of a pervasive IoT can be achieved, there is a requirement to invest in the infrastructure necessary to integrate constrained devices into the fog hierarchy.
This idea is similar to what is proposed by mist computing, but we emphasized that constrained devices connected to the fog do not necessarily need to be specialized nor dedicated, as there is much potential in allowing them to receive and execute custom user code instead of sending the data to more powerful fog nodes.

Furthermore, we organized the classification of constrained devices presented by the IETF and used it to categorize current devices.
It is possible to notice that popular gadgets such as the Arduino Uno Rev3, the \nth{2}-generation Nest Thermostat, and the Fitbit activity tracker are all considered constrained devices.
This means that, although there have been several technological advancements in the last decades, constrained devices still play an important role today and their importance can be expected to grow along with the wide adoption of the IoT.

We also categorized applications according to their resource requirements, identifying and giving examples for three different classes.
Namely, a class of applications well-suited for constrained devices (e.g., filters), one for applications that demand a lot of resources, but can make several constrained devices cooperate with each other to execute them (e.g., deep learning), and one for applications that are incompatible with constrained devices (e.g., high-resolution video streaming processing).

By providing both hardware and software classifications, we intend to help developers assess the characteristics of the problem with which they are working, as this can be useful in the process of implementing a solution for it.

We then discussed our expectations for a scenario that takes place a few decades from now, where a great number of objects connected to the Internet will be very small devices and depicted use cases that can be seen as the next step for current and near-future IoT applications.
In this future, having sensors embedded in items as trivial as candy wrappers may be an everyday reality.

Finally, we presented several challenges and opportunities that will arise in this path to enabling the full potential of the IoT, and we invite the reader to join us in the exploration of this topic, as many possibilities are waiting to be uncovered.

\section*{Acknowledgment}
The authors would like to thank CAPES and CNPq for the financial support. 
This work was partially supported by grants 2018/02204-6 and \mbox{2015/24494-8} from FAPESP and it is part of the INCT project called the Future Internet for Smart Cities (CNPq 465446/2014-0, CAPES 88887.136422/2017-00, and FAPESP 2014/50937-1 and 2015/24485-9). 
E. Borin is partially funded by CNPq (313012/2017-2), FAPESP (2013/08293-7), and Petrobras.

\newpage

\section*{Acronyms and Abbreviations}
\label{sec:acronyms_abbreviations}
\begin{itemize}[label={}]
\setlength\itemsep{0.5em}
\item \textbf{3D} Three-dimensional
\item \textbf{6LoWPAN} IPv6 over Low-Power Wireless Personal Area Networks
\item \textbf{API} Application Programming Interface
\item \textbf{B} Byte (unit of memory size)
\item \textbf{CNN} Convolutional Neural Network
\item \textbf{CoAP} Constrained Application Protocol
\item \textbf{CPU} Central Processing Unit
\item \textbf{FPU} Floating-Point Unit
\item \textbf{GPU} Graphics Processing Unit
\item \textbf{Hz} Hertz (unit of frequency)
\item \textbf{HTTP} Hypertext Transfer Protocol
\item \textbf{IEC} International Electrotechnical Commission
\item \textbf{IEEE} Institute of Electrical and Electronics Engineers
\item \textbf{IETF} Internet Engineering Task Force
\item \textbf{IoT} Internet of Things
\item \textbf{IP} Internet Protocol
\item \textbf{kbps} Kilobits per second (multiple of the unit of data transfer rate)
\item \textbf{KiB} Kibibytes (IEC multiple of the unit of memory size)
\item \textbf{LAN} Local Area Network
\item \textbf{MAC} Media Access Control (OSI model)
\item \textbf{MCC} Mobile Cloud Computing
\item \textbf{MCU} Microcontroller Unit
\item \textbf{MEC} Multi-access Edge Computing
\item \textbf{MiB} Mebibytes (IEC multiple of the unit of memory size)
\item \textbf{NIST} National Institute of Standards and Technologies
\item \textbf{NFC} Near-Field Communication
\item \textbf{PHY} Physical Layer (OSI model)
\item \textbf{QoS} Quality of Service
\item \textbf{RAM} Random Access Memory
\item \textbf{RAN} Radio Access Network
\item \textbf{RFID} Radio-Frequency Identification
\item \textbf{RISC} Reduced Instruction Set Computer
\item \textbf{RPL} Routing Protocol for Low Power and Lossy Networks
\item \textbf{TLS} Transport Layer Security
\item \textbf{UDP} User Datagram Protocol
\item \textbf{USA} United States of America
\item \textbf{VM} Virtual Machine
\item \textbf{WSAN} Wireless Sensor and Actuator Networks
\item \textbf{WSN} Wireless Sensor Network
\end{itemize}

\section*{Glossary}
\label{sec:glossary}
\begin{itemize}[label={}]
\setlength\itemsep{0.7em}
\item \textbf{Cloud computing:} model that aims at enabling access to a shared pool of computational resources, such as networks, servers, storage, applications, and services~\cite{Mell2011-NIST_Cloud}.

\item \textbf{Constrained device:} device that does not have some of the characteristics (e.g., power, memory, or processing resources) that are expected to be present on other devices currently connected to the Internet~\cite{Bormann2014-TerminologyConstrained}.

\item \textbf{Fog computing:} paradigm that aims at filling a gap between cloud computing data centers and end-devices~\cite{Chiang2017-10questions}, a layered model for enabling ubiquitous access to a shared continuum of scalable computing resources~\cite{NIST2018-FogComputingConceptualModel}.

\item \textbf{Internet of Things:} dynamic network infrastructure with self-configuring capabilities based on standards, its components are physical or virtual ``things'' that have attributes as well as identities and are able to use intelligent interfaces and be part of an information network~\cite{Li2015-IoTSurvey}.

\item \textbf{Mist computing:} lightweight and rudimentary form of fog computing that exists at the edge of the network, it uses microcomputers and microcontrollers to feed into fog computing nodes and potentially onward toward centralized cloud computing services~\cite{NIST2018-FogComputingConceptualModel}.

\item \textbf{Resource-constrained device:} See constrained device.

\end{itemize}

\newpage

\bibliographystyle{unsrt} 

\bibliography{ios-book-article}

\section*{Author Biographies}
\label{sec:author_biographies}

{
\setlength\parindent{0pt}

\parbox[t][1.6in][t]{\textwidth}{
\begin{wrapfigure}{l}{20mm}
    \vspace{-13pt}
    \includegraphics[width=1in,clip,keepaspectratio]{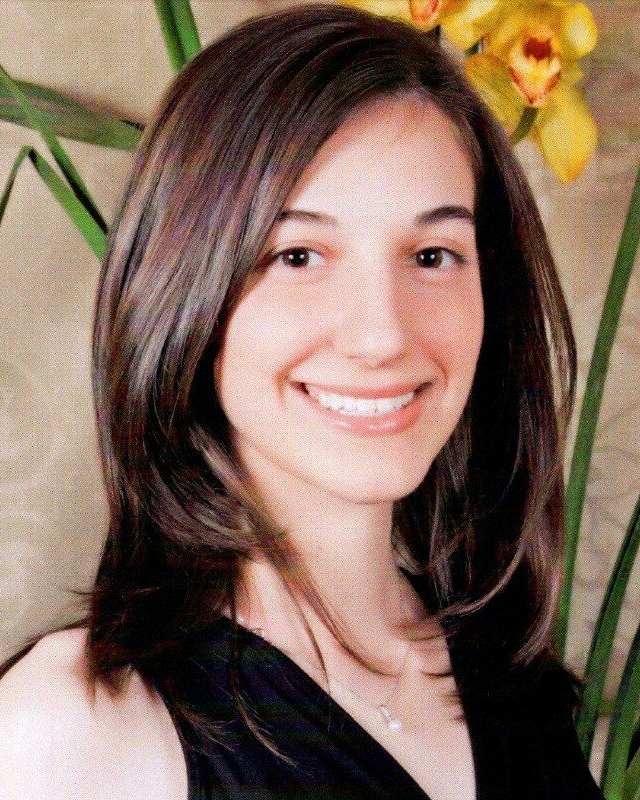}
\end{wrapfigure}\par
\textbf{Fl{\'{a}}via Pisani} received a B.Sc. degree with honors in Computer Science and a Ph.D. in Computer Science from the University of Campinas (UNICAMP) in 2014 and 2019, respectively.
She is currently a postdoctoral researcher at the Department of Informatics of the Pontifical Catholic University of Rio de Janeiro (PUC-Rio).
Her current research interests include high-performance computing, parallel computing, fog computing, constrained devices, the Internet of Things, smart cities, Computer Science education, computational thinking, and the use of technology in education.\par
}

\vspace{1.5cm}

\parbox[t][1.425in][t]{\textwidth}{
\begin{wrapfigure}{l}{20mm}
    \vspace{-13pt}
    \includegraphics[width=1in,clip,keepaspectratio]{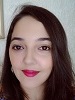}
\end{wrapfigure}\par
\textbf{Fab{\'{i}}ola M. C. de Oliveira} received a B.Sc. in Control and Automation Engineering from the University of Campinas (Brazil) in 2012 and an M.Sc. in Mechanical Engineering from the same university in 2015.
Currently, she is a Ph.D. candidate at the Institute of Computing, University of Campinas.
Her research interests include graph partitioning, deep learning distributed applications, fog computing, Internet of Things, distributed computing, parallel computing, and high-performance computing.\par
}

\vspace{1.5cm}

\parbox[t][1.425in][t]{\textwidth}{
\begin{wrapfigure}{l}{20mm}
    \vspace{-13pt}
    \includegraphics[width=1in,clip,keepaspectratio]{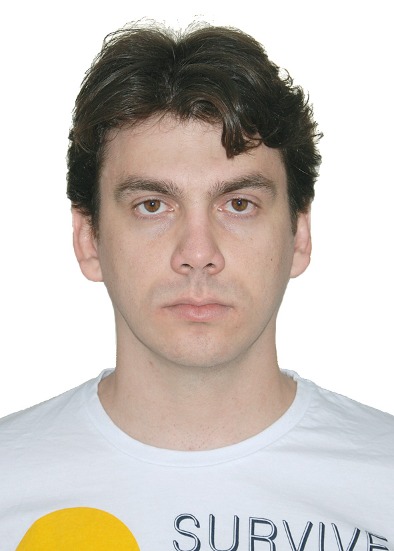}
\end{wrapfigure}\par
\textbf{Eduardo S. Gama} received a M.Sc. degree and a B.Sc. degree in Computer Science from the Federal University of Paraiba~(UFPB) in 2015 and 2017, respectively. Currently, he is a Ph.D. student in Computer Science at the Institute of Computing from the University of Campinas~(UNICAMP). His current research interests are edge computing, cloud computing, multimedia networks, Quality of Experience for edge/cloud computing.\par
}

\vspace{1.5cm}

\parbox[t][1.6in][t]{\textwidth}{
\begin{wrapfigure}{l}{20mm}
    \vspace{-13pt}
    \includegraphics[width=1in,clip,keepaspectratio]{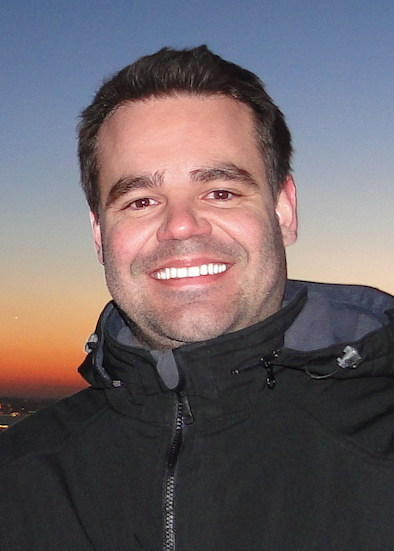}
\end{wrapfigure}\par
\textbf{Roger Immich} is an Assistant Professor at the Digital Metropolis Institute (IMD) of the Federal University of Rio Grande do Norte (UFRN). He received his Ph.D. in Informatics Engineering from the University of Coimbra, Portugal (2017). He was a visiting researcher at the University of California at Los Angeles, United States (UCLA) in 2016/2017 and a postdoctoral researcher at the Institute of Computing of the University of Campinas (UNICAMP) in 2018/2019. His research interests are Smart Cities, IoT, 5G, Quality of Experience, as well as Cloud and Fog computing.\par
}

\vspace{1.5cm}

\parbox[t][1.95in][t]{\textwidth}{
\begin{wrapfigure}{l}{20mm}
    \vspace{-13pt}
    \includegraphics[width=1in,clip,keepaspectratio]{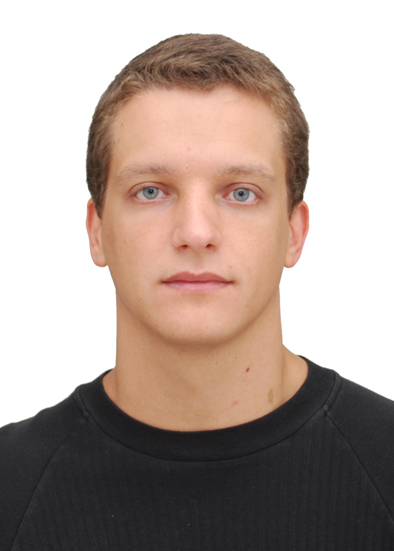}
\end{wrapfigure}\par
\textbf{Luiz F. Bittencourt} is an Associate Professor at the University of Campinas (UNICAMP), Brazil. He was a visiting researcher in the University of Manchester, UK, Cardiff University, UK, and Rutgers University, USA. Luiz was awarded the IEEE Communications Society Latin America Young Professional Award in 2013. He acts in the organization of several conferences in cloud computing and edge computing subjects. He also serves as associate editor for the IEEE Cloud Computing Magazine, for the Computers and Electrical Engineering journal, and for the Internet of Things Journal. His main interests are topics related to resource management and scheduling in cloud and fog computing.\par
}

\vspace{1.5cm}

\parbox[t][1.6in][t]{\textwidth}{
\begin{wrapfigure}{l}{20mm}
    \vspace{-13pt}
    \includegraphics[width=1in,clip,keepaspectratio]{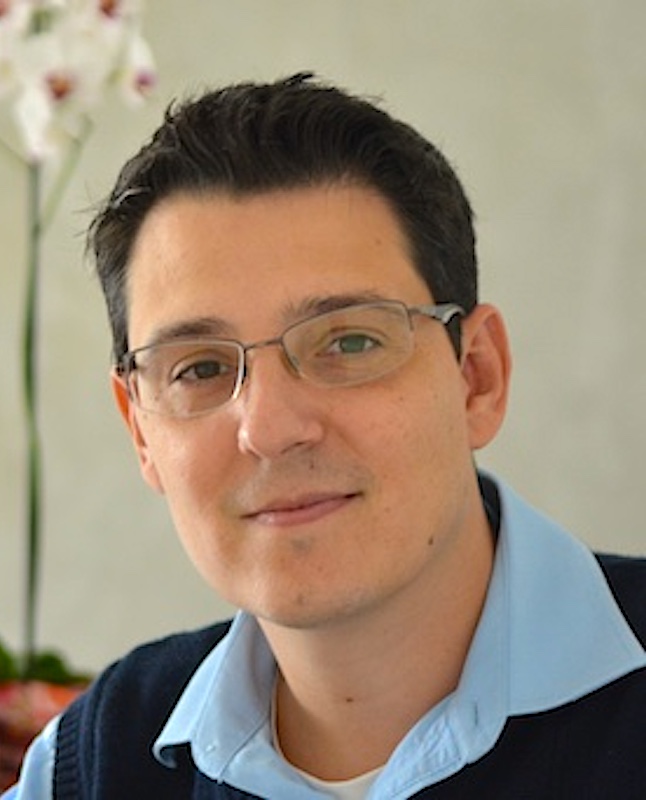}
\end{wrapfigure}\par
\textbf{Edson Borin} received a Ph.D. degree in Computer Science from the University of Campinas (Brazil) in 2007 and joined Intel Labs, where he developed research on compilers, microarchitectural, and architectural techniques to accelerate software on modern microprocessors. Since 2010, he has been a faculty member at the Institute of Computing, University of Campinas (UNICAMP) and his research interests include compilers, computer architecture, programming models, parallel and distributed systems, and high-performance computing.\par
}

}
\end{document}